\documentclass[final,5p,times,twocolumn]{elsarticle}

\usepackage{amssymb}
\usepackage{chemformula}
\usepackage{siunitx}
\usepackage[percent]{overpic}
\usepackage{pgfplots} 
\usepackage{caption}
\usepackage{subcaption}
\usepackage{amsfonts}
\usepackage{multirow}
\usepackage{bm}
\usepackage{tikz}
\usetikzlibrary{shapes.geometric}
\usepackage{epstopdf}

\journal{Acta Astronautica}

\begin{document}

\begin{frontmatter}



\title{Satellite design optimization for differential lift and drag applications}


\author[inst1]{C. Marianowski}
\author[inst1]{C. Traub}
\author[inst1]{M. Pfeiffer}
\author[inst1]{J. Beyer}
\author[inst1]{S. Fasoulas}

\affiliation[inst1]{organization={Institute of Space Systems, University of Stuttgart},
            addressline={Pfaffenwaldring 29}, 
            city={Stuttgart},
            postcode={70569},
            country={Germany}}

\begin{abstract}
Utilizing differential atmospheric forces in the Very Low Earth Orbits (VLEO) regime for the control of the relative motion within a satellite formation is a promising option as any thrusting device has tremendous effects on the mission capacity due to the limited weight and size restrictions of small satellites. One possible approach to increase the available control forces is to reduce the mass of the respective satellites as well as to increase the available surface area. However, satellites of these characteristics suffer from rapid orbital decay and consequently have a reduced service lifetime. Therefore, achieving higher control forces is in contradiction to achieving a minimum orbital decay of the satellites, which currently represents one of the biggest challenges in the VLEO regime. In this work, the geometry of a given reference satellite, a 3UCubeSat, is optimized under the consideration of different surface material properties for differential lift and drag control applications while simultaneously ensuring a sustained VLEO operation. Notably, not only the consideration of sustainability but also the optimization with regard to differential lift is new in literature. It was shown that the advantageous geometries strongly depend on the type of gas-surface interaction and thus, two different final designs, one for each extreme type, are presented. In both cases, improvements in all relevant parameters could be achieved solely via geometry adaptions.

\end{abstract}



\begin{keyword}
VLEO \sep satellite aerodynamics \sep shape optimization \sep differential drag and lift
\end{keyword}

\end{frontmatter}


\section{Introduction}
\label{sec:Introduction}
In the recent past, interest in operating satellites at much lower altitudes than before, in the so-called VLEO regime (i.e. the entirety of orbits with a mean altitude $<450$~km \cite{Crisp.2020}), has increased due to a variety of advantages, which have been summarized by Crisp et al. \cite{Crisp.2020}. Similarly, the application of distributed satellite systems, which commonly are made up of several small satellites working together to achieve a shared goal, is nowadays prevalent \cite{damico}. Due to their stringent volume and mass limitations, alternative solutions to the conventional propulsion methods to exert control forces are of great interest. In the VLEO regime, the use of the atmospheric forces from the prevailing residual atmosphere represents a promising solution.

At the Institute of Space Systems of the University of Stuttgart, this methodology is actively researched since 2018 \cite{traub2020exploitation}. In the most recent publication, a planning tool for optimal three-dimensional formation flight maneuvers of satellites in VLEO using aerodynamic lift and drag via yaw angle deviations has been presented \cite{traub2022planning}. As in the VLEO regime the large levels of orbital decay represent the major challenge to be overcome for a sustained operation to become reality, the planned trajectory is optimal in a sense that the overall decay during the maneuver is minimized.
Thereby, the remaining lifetime of the satellites is maximized and the practicability and sustainability of the methodology significantly increased. Additionally, applying yaw angle deviations allows to simultaneously control the in- and out-of-plane relative motion via differential lift and drag. Throughout the work of Traub et al. \cite{traub2022planning}, however, conventional 3UCubeSats augmented with two solar panels have been assessed. In parallel efforts, optimized shapes for VLEO satellites targeting a minimization
of the atmospheric drag force and thus extension of operational lifetime have been developed. Thereby, the satellite geometry has been optimized via a novel 2D profile optimization tool \cite{Hild.2022}. It was shown that the optimized satellite geometries offer pure passive lifetime extensions of up to 46~\% compared to a GOCE like reference body.

In this article, which is based on the corresponding author's master's thesis \cite{Marianowski.2022} and builds upon said contributions, optimal satellite designs for differential lift and drag applications are presented. To the best of the authors' knowledge, such efforts have never been discussed in the literature. The article elaborates how the research objective leads to conflicting requirements, which
need to be prioritised accordingly to arrive at an overall optimal design. All optimization steps are presented in detail and the corresponding considerations are discussed. In future efforts, it is foreseen to combine the optimal maneuver planning and the optimal satellite designs, which would represent a more holistic optimization approach.

The necessary fundamentals are briefly presented in Section~\ref{sec:Satellite_Aerodynamics}. In Section~\ref{sec:optimization_Approach}, the for the calculations and optimization set parameters and constraints as well as the applied optimization steps are introduced. The design optimization for diffuse re-emission is conducted in Section~\ref{sec:process1}, and the similar process for specular reflection can be found in Section~\ref{sec:process2}.
\section{Satellite aerodynamics}
\label{sec:Satellite_Aerodynamics}
\subsection{Fundamentals}
The forces acting on a satellite are the result of the interchange of momentum between
the particles of the prevailing atmosphere and the satellite. Whereas the
atmospheric forces attacking at the center of gravity are of interest within this work, the torque induced by the atmosphere was neglected and consigned to the attitude control system of the satellite. The specific drag force $\vec{f}_{D}$ and specific lift force $\vec{f}_{L}$ for a satellite of mass $m$ can be calculated as shown in Eq.~\ref{eq:ad} and Eq.~\ref{eq:al}:
\begin{equation}
\label{eq:ad}
\vec{f}_{D}=\frac{1}{2}\cdot \rho \cdot \frac{C_D \cdot A_{ref}}{m} \cdot |\vec{v}_{rel}|^2 \cdot \frac{\vec{v}_{rel}}{|\vec{v}_{rel}|},
\end{equation}
\begin{equation}
\label{eq:al}
\vec{f}_{L}=\frac{1}{2}\cdot \rho \cdot \frac{C_L \cdot A_{ref}}{m} \cdot |\vec{v}_{rel}|^2 \cdot \frac{(\vec{v}_{rel}\times \vec{n})\times \vec{v}_{rel}}{|(\vec{v}_{rel}\times \vec{n})\times \vec{v}_{rel}|},
\end{equation}
where $\rho$ is the prevailing density, $C_D$ the drag coefficient, $C_L$ the lift coefficient, $A_{ref}$ the reference area, which corresponds to the projected area of the satellite perpendicular to the flow, and $\vec{v}_{rel}$ the relative velocity to the local atmosphere. In Eq.~\ref{eq:al}, $\vec{n}$ denotes the surface normal vector of the surface under consideration.

In order to investigate the sustainability of the different satellite designs, their lifetime was estimated. To ensure comparability with the previous study \cite{Hild.2022}, the same simplified equation for the lifetime of a satellite in a circular orbit was applied \cite{Walter.2019}:
\begin{equation}
\label{eq:tl}
t_l = \frac{\beta \cdot H_0}{\rho_0 \cdot \sqrt{\mu_E \cdot a}} \cdot \left( 1-exp \left( -\frac{h_0}{H_0} \right) \cdot \left( 1+\frac{h_0}{2a} \right)\right).
\end{equation}
Here, $t_l$ is the lifetime of the satellite for a given orientation and no additional use of propulsion. $H_0$ is the atmospheric scale height, $h_0$ the base altitude and $\rho_0$ the density as shown in Tab.~\ref{tab:piecewise_exponential_model}. $\mu_E$ is the Earth's gravitational parameter and $a=h_0 + R_E$ with $R_E$ being the Earth's radius.
\begin{table}[tb]
	\centering
		\begin{tabular}{c|ccc}  
		$h$ [\si{\kilo\metre}] & $h_0$ [\si{\kilo\metre}] & $\rho_0$ [\si{\kilo\gram\per\cubic\metre}] & $H_0$ [\si{\kilo\metre}]\\ \hline 
		250-300 & 250 & 7.248$\cdot$10\textsuperscript{-11} & 45.546\\ 
		300-350 & 300 & 2.418$\cdot$10\textsuperscript{-11} & 53.628\\
		350-400 & 350 & 9.518$\cdot$10\textsuperscript{-12} & 53.298\\
		400-450 & 400 & 3.725$\cdot$10\textsuperscript{-12} & 58.515\\ 
	\end{tabular}
	\caption{Parameters for Eq.~\ref{eq:tl} \cite{Walter.2019}.}
	\label{tab:piecewise_exponential_model}%
\end{table}
Although Eq.~\ref{eq:tl} is very simplified due to the neglected changes in density with time and location, it is well suited for the purpose of this work, since all atmospheric changes have to be eliminated to only account for the influence of the satellite's design.
\subsection{Aerodynamic coefficients and gas-surface-interaction}
\label{sec:Aerodynamic_Coefficients}
\begin{figure}
		\centering
		\begin{overpic}[width=0.47\textwidth]{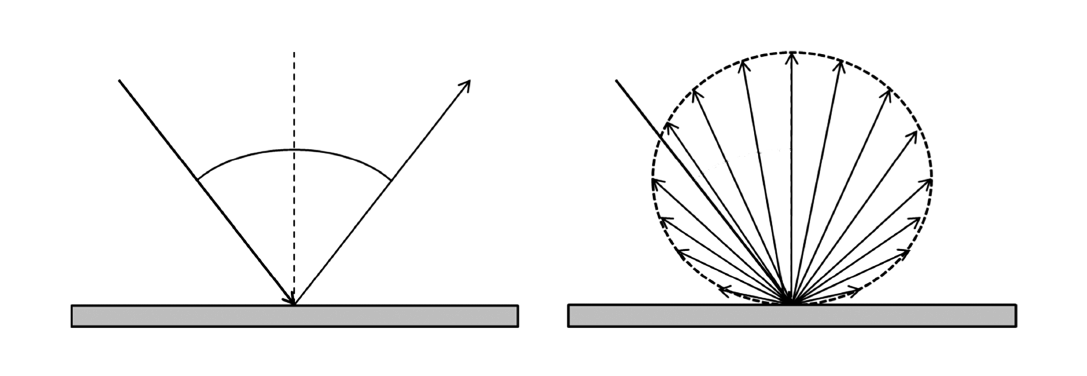}
    		\put (2,28) {\scriptsize Incident Flux}
    		\put (48,28) {\scriptsize Incident Flux}
    	\end{overpic}
		\caption{Specular reflection (left) and diffuse re-emission (right) \cite{Walker.2014}.}
		\label{subfig:maxwell}
	\end{figure}

In Eq.~\ref{eq:ad} and Eq.~\ref{eq:al}, the parameters $m$, $A_{ref}$, $C_D$, and $C_L$ are characteristics of the satellite's design. Whereas the first two can be directly derived and measured using the satellite or its model, the two aerodynamic coefficients are depending on the surface properties of the satellite and have to be calculated analytically or numerically. The two main parameters influencing the aerodynamic coefficients are the velocity of the reflected particle, which is depending on the amount of exchanged energy between particle and wall, commonly described using energy accommodation coefficients, and the angle of reflection/re-emission, which is additionally depending on the type of gas-surface-interaction (GSI). There are two extreme types of GSI; specular reflections, where the angle between surface and reflected particle is equal to the angle between surface and incoming particle, and diffuse re-emissions, with a particle re-emission according to a probabilistic velocity and direction distribution (see Fig.~\ref{subfig:maxwell}). 

In general, specularly reflecting material are advantageous as they allow a deliberate deflection of the incoming particles. However, specularly reflecting materials are not yet available and remain subject of current research \cite{roberts, crisp.2021}. Hence, the prevailing type of GSI in VLEO on the currently used materials is a diffuse or quasi-diffuse gas-surface interaction \cite{Moe, Pilinski}. 
\subsubsection{Accommodation coefficients}
\label{subsec:accommodation_coefficients}
The thermal energy accommodation coefficient 
\begin{equation}
\label{eq:at}
\alpha_T = \frac{E_i - E_r}{E_i - E_w}
\end{equation}
is a measure for the adaption of energy from the impinging particles $E_i$ to the energy $E_w$, which the particles would have after reaching equilibrium with the satellite's wall temperature. $E_r$ corresponds to the actual energy of the reflected particles. Similarly, the momentum exchange between particle and surface is commonly described by the tangential momentum accommodation coefficient $\sigma_t$ and the normal momentum accommodation coefficient $\sigma_n$ as follows \cite{Chambre.1958}:
\begin{equation}
\label{eq:sigma_t}
\sigma_t = \frac{\tau_i-\tau_r}{\tau_i-\tau_w},
\end{equation}
\begin{equation}
\label{eq:sigma_n}
\sigma_n = \frac{p_i-p_r}{p_i-p_w}.
\end{equation}
Here, $\tau_i$ is the tangential momentum carried to the surface by the incident particle and $\tau_r$ is the tangential momentum carried away from the surface by the reflected particle. The tangential momentum carried away from the surface by a diffusely reflected particle after reaching thermal equilibrium with the wall is ascribed to $\tau_w$ and is per definition equal to zero. The normal momentum carried to the surface by the incident particle and normal momentum carried away from the surface by the reflected particle are $p_i$ and $p_r$ respectively, and $p_w$ is the normal momentum carried away from the surface by a diffusely reflected particle after reaching thermal equilibrium with the wall.

\subsection{Differential lift and drag control method}
\label{sec:Differnetial_Lift}
In order to change the formation geometry, the method of differential lift and drag is generally understood to intentionally create differences in the acting aerodynamic forces  \cite{traub2020exploitation}, e.g. by rotating only one of the two satellites of the formation (see Fig.~\ref{fig:diff_drag}). Whereas differential drag effects the relative movement within the orbital plane, differential lift effects the out-of-plane relative movement. The maximum achievable differential specific forces are 
\begin{equation}
\label{eq:admax}
    |\vec{f}_{\Delta D}|=|\vec{f}_{D,max}|-|\vec{f}_{D,min}|,
\end{equation}
\begin{equation}
\label{eq:almax}
    |\vec{f}_{\Delta L}|=|\vec{f}_{L,max,1}| + |\vec{f}_{L,max,2}|=2 \cdot |\vec{f}_{L,max}|,
\end{equation}
if $|\vec{f}_{L,max,1}|=|\vec{f}_{L,max,2}|$ and $\vec{f}_{L,max,1} \uparrow \downarrow \vec{f}_{L,max,2}$ as shown in Fig.~\ref{subfig:lift_configuration}.

\begin{figure}[tb]
	\centering
	\begin{overpic}[width=0.47\textwidth]{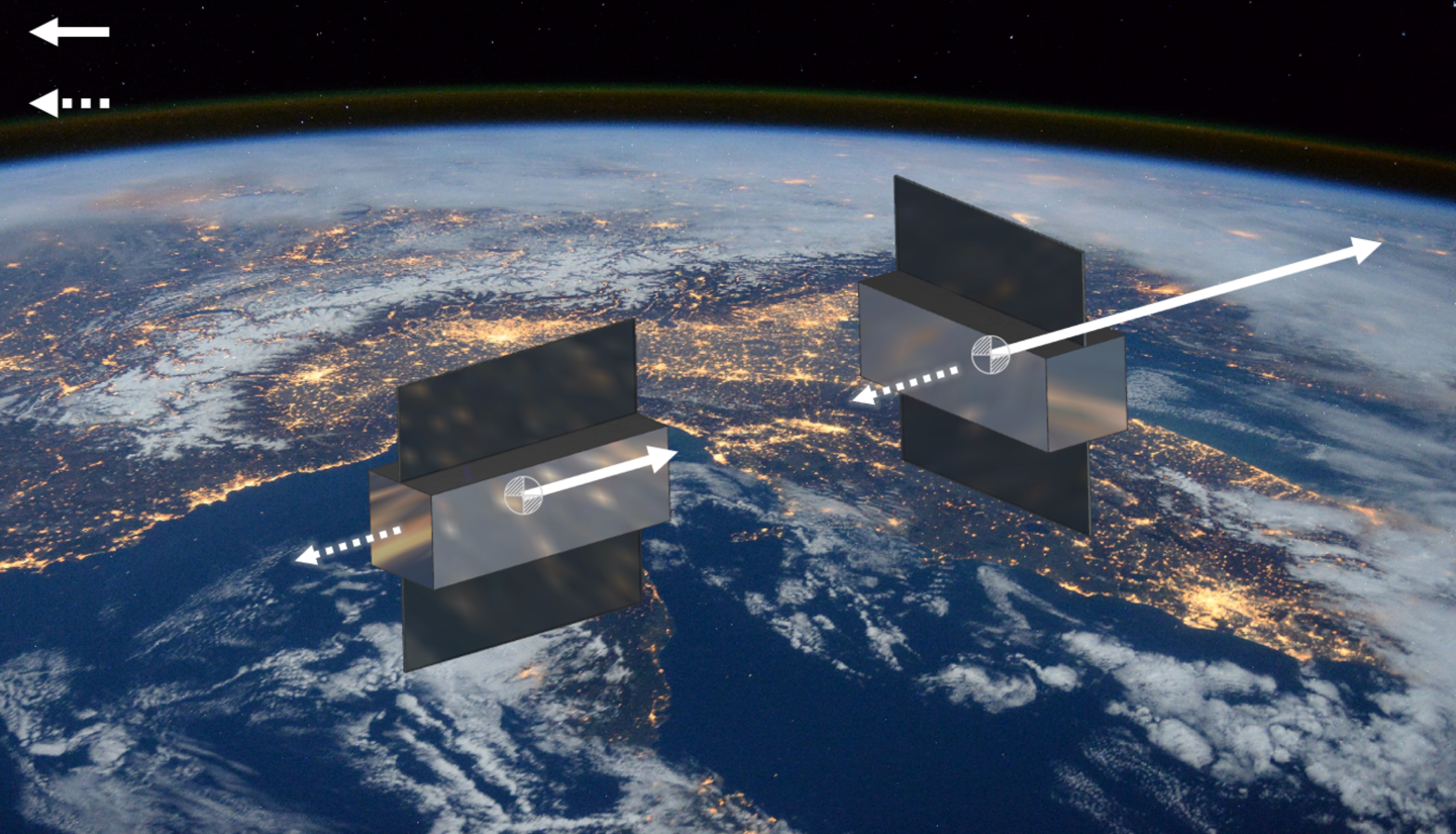}
		\put (8.5,49.4){\textcolor{white} {\small Flight Direction}}
		\put (8.5,54.2){\textcolor{white} {\small Drag Force}}
	\end{overpic}
	\caption{Depiction of a maximum differential drag force.}
	\label{fig:diff_drag}
\end{figure}
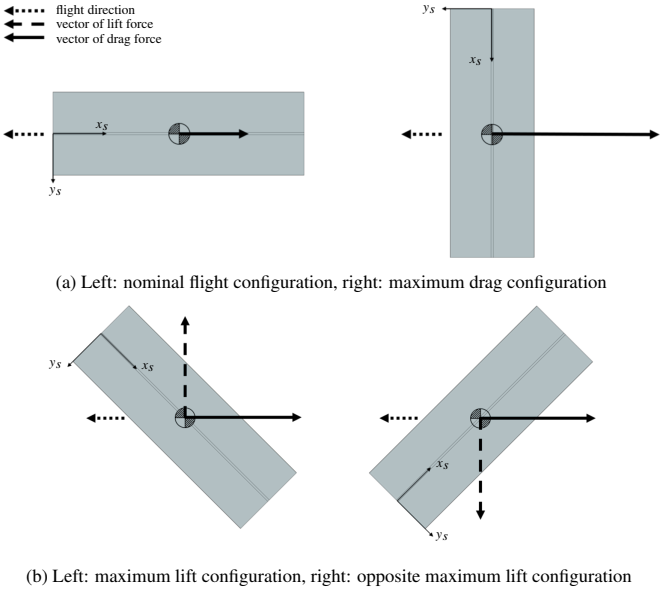
\begin{figure}
\centering
\valign{#\cr
  \hbox{\subfloat[Left: nominal flight configuration, right: maximum drag configuration\label{subfig:drag_configuration}]{%
		\begin{overpic}[width=0.47\textwidth]{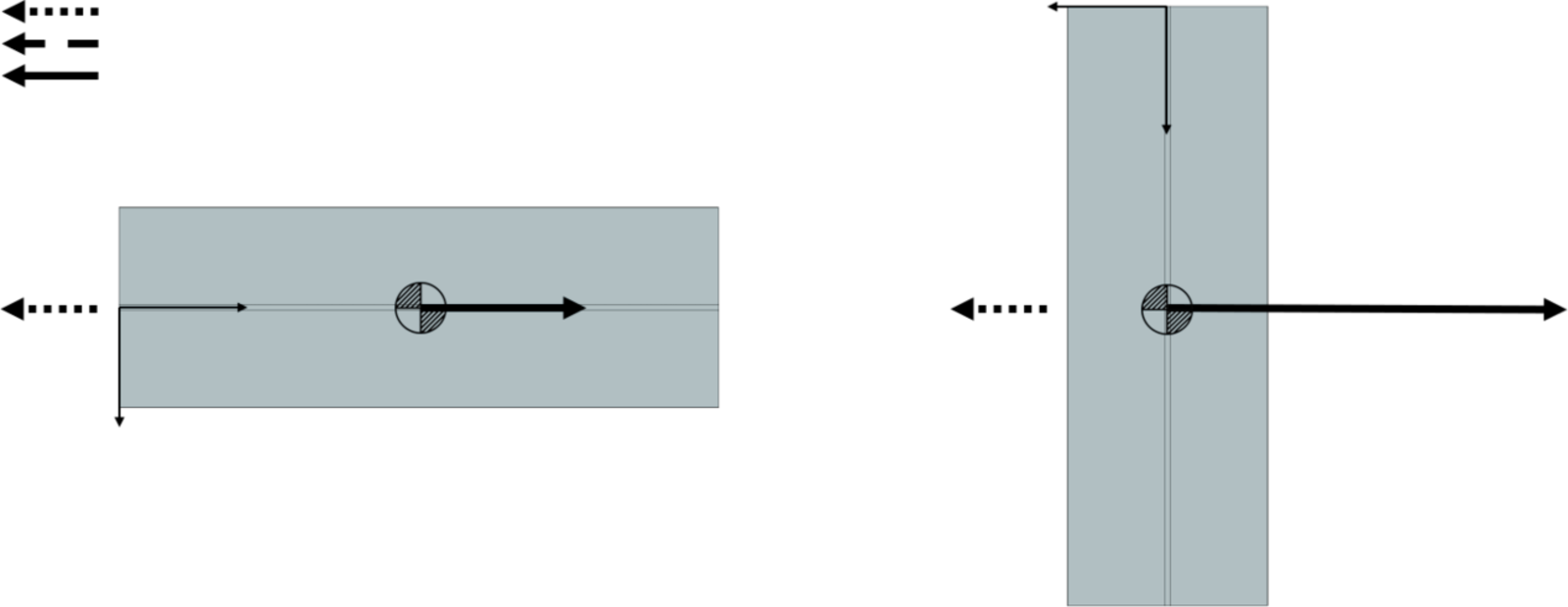}
    		\put (8,37.5) {\tiny flight direction}
    		\put (8,35.25) {\tiny vector of lift force}
    		\put (8,33) {\tiny vector of drag force}
    		\put (14,20){\tiny $x_s$}
    		\put (7,10){\tiny $y_s$}
    		\put (71,30){\tiny $x_s$}
    		\put (64,38){\tiny $y_s$}
    	\end{overpic}}
  }
  \hbox{\subfloat[Left: maximum lift configuration, right: opposite maximum lift configuration \label{subfig:lift_configuration}]{%
		\begin{overpic}[width=0.47\textwidth]{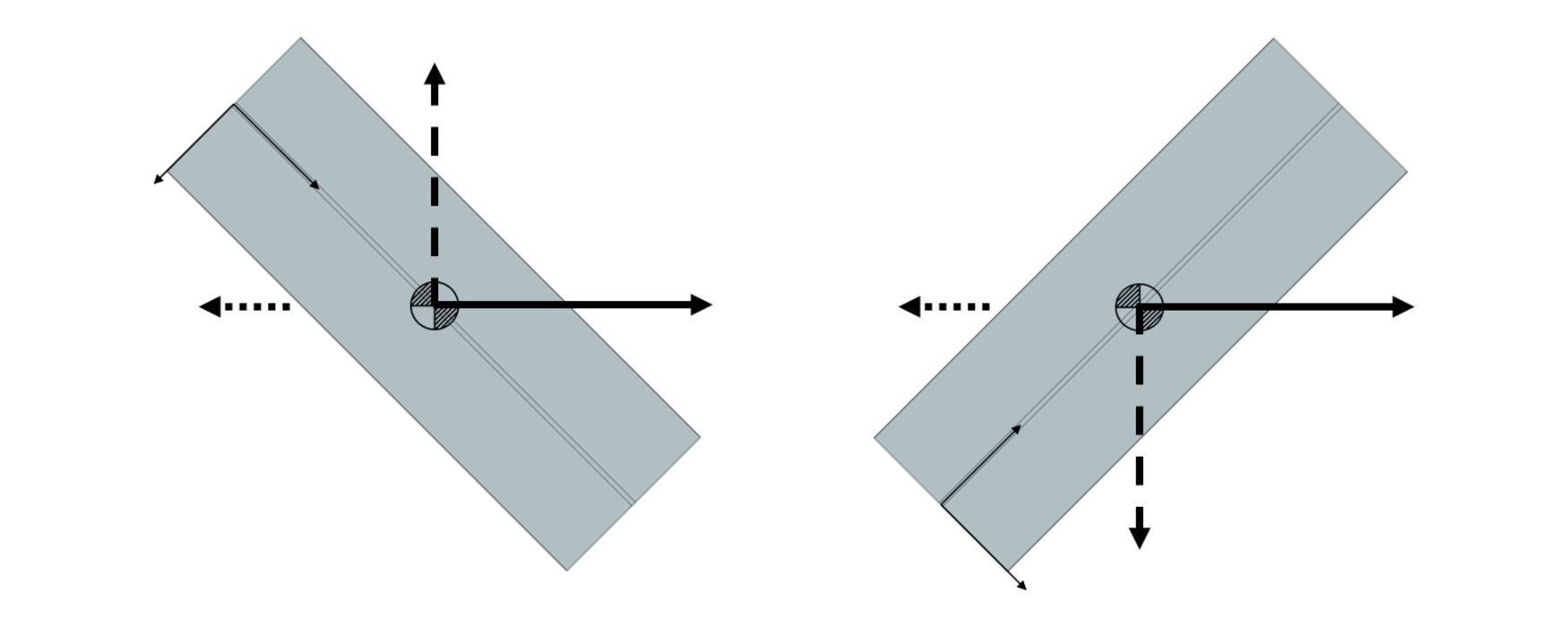}
			\put (21,28){\tiny $x_s$}
    		\put (7,28.5){\tiny $y_s$}
    		\put (66,13){\tiny $x_s$}
    		\put (66,2){\tiny $y_s$}
    	\end{overpic}}
  }\cr
}
\caption{Depiction of the different possible orientations of a satellite depending on the desired control forces. Here: reference satellite (Fig.~\ref{fig:default_satellite}) in top view.}
\label{fig:flight_configurations}
\end{figure}

Within this work, it is assumed that all satellites within the formation share the same design. The following flight configurations\footnote{In this article, the term \textit{"(flight) configuration"} is used to describe the orientation of a satellite, which can be further defined by the angle between the satellite's longitudinal axis and the orbital plane (angle of sideslip - AOS) as well as the angle between the longitudinal axis and the plane formed by $\vec{v}_{rel}$ and the orbit's normal vector (angle of attack - AOA).} are defined, of which Fig.~\ref{fig:flight_configurations} gives an overview of:
\begin{itemize}
\item A nominal flight configuration experiencing as little drag as possible and no/negligible lift ($\vec{f}_{D,min}$, see Fig.~\ref{subfig:drag_configuration} left),
\item A maximum drag configuration for experiencing the highest possible drag ($\vec{f}_{D,max}$, see Fig.~\ref{subfig:drag_configuration} right),
\item A maximum lift configuration for experiencing the highest possible lift ($\vec{f}_{L,max}$, see Fig.~\ref{subfig:lift_configuration}).
\end{itemize}

 With the aim of optimizing the satellite's design for a sustainable use of differential lift and drag control methods in VLEO, the specific optimization goals  can be expressed as follows:
\begin{itemize}
\item Drag Goal \#1: Decreasing $\vec{f}_{D,min}$
\item Drag Goal \#2: Increasing $\vec{f}_{D,max}$
\item Lift Goal \#1: Increasing $\vec{f}_{L,max}$
\end{itemize}

In order to derive design recommendations, the different design characteristics are prioritized as they contribute to contradictory optimization goals. Considering that aerodynamic force control methods are not the means of choice for a time-critical maneuver due to the comparatively low absolute values of specific forces, the overall priority is to decrease the orbital decay. It can be assumed that the satellite spends most of its service life in the nominal flight configuration, and hence as little unwanted aerodynamic forces as possible are desired. Due to the prevailing condition of the contamination with atomic oxygen on currently used materials and a therefore diffuse gas-surface-interaction with generally little $C_L$ values, increasing the lift force is set to be the second priority within this work.

To summarize, the priorities underlying all decisions within this work are set as follows:
\begin{equation*}
\text{Priorities:} \quad \uparrow t_l \quad > \quad \uparrow \vec{f}_{\Delta L} \quad > \quad \uparrow \vec{f}_{\Delta D},
\end{equation*} 
with $\uparrow$ indicating an increase. However, these priorities are based on the considerations mentioned above, and naturally may differ for individual missions.
\section{Optimization approach}
\label{sec:optimization_Approach}
The framework conditions set to pursue the defined goals alongside the utilized tools and optimization steps taken are presented in the following sub-sections.

\subsection{Boundary conditions of the optimization}
In this section, the chosen atmospheric conditions for the determination of the forces, as well as the chosen reference satellite and the derived geometry constraints for the design variations are briefly presented.
\subsubsection{Atmospheric conditions}
The atmospheric conditions are set constant for all considerations within this work in order to ascribe changes in the specific forces to the design variations. Therefore, moderate space weather with a 10.7~cm flux $F_{10.7}$ of 140 sfu and a geomagnetic activity index of $A_p = 15$ were chosen\footnote{Detailed information about solar proxies and indices for space weather description can be found in the work of Doornbos \cite{Doornbos.2012}.}. The data shown in Tab.~\ref{tab:Atmosphere_Data} was obtained by averaging the output of the NRLMSISE-00 model \cite{Picone.2002} of one day per month of 2004 over the year for the chosen altitude of 350 km. Additionally, the satellite's wall temperature is set to $T_w=300$~\si{\kelvin} as commonly assumed in the regarding literature \cite{Doornbos.2012, llop}.
\begin{table}[tb]
	\centering
		\begin{tabular}{c|l|l}
        Parameter & Unit & Value \\ \hline
		$T_i$  & \si{\kelvin}& 1,056.6  \\ 
		$\rho$  &  \si{\kilo\gram\per\cubic\metre}& 9.15 $\cdot$10\textsuperscript{-12}  \\ 
		$M$ & \si{\kilo\gram\per\mol}& 0.0174  \\ 
		$v_{rel}$ & \si{\meter\per\second}& 7,697.1  \\
		$s$ & - & 7.66 \\
		$n_O$ & \si{\per\cubic\metre}& 2.64$\cdot$10\textsuperscript{14}  \\
		$n_{N2}$ & \si{\per\cubic\metre}& 4.18$\cdot$10\textsuperscript{13}  \\
		$n_{He}$ & \si{\per\cubic\metre}& 4.88$\cdot$10\textsuperscript{12}  \\
		$n_N$ & \si{\per\cubic\metre}& 4.44$\cdot$10\textsuperscript{12}  \\
		$n_{O2}$ & \si{\per\cubic\metre}& 1.09$\cdot$10\textsuperscript{12}  \\
		$n_H$ & \si{\per\cubic\metre} &8.53$\cdot$10\textsuperscript{10} \\
		$n_{Ar}$ & \si{\per\cubic\metre}&8.63$\cdot$10\textsuperscript{9}  \\ 
	\end{tabular}
	\caption{Atmospheric data used for all calculations in this work: temperature of the impinging particle $T_i$, molecular mass $M$, molecular speed ratio $s$, and particle density $n_i$ of species i.}
	\label{tab:Atmosphere_Data}%
\end{table}
\subsubsection{Reference satellite}
\begin{figure}
\centering
	\begin{overpic}[width=0.35\textwidth]{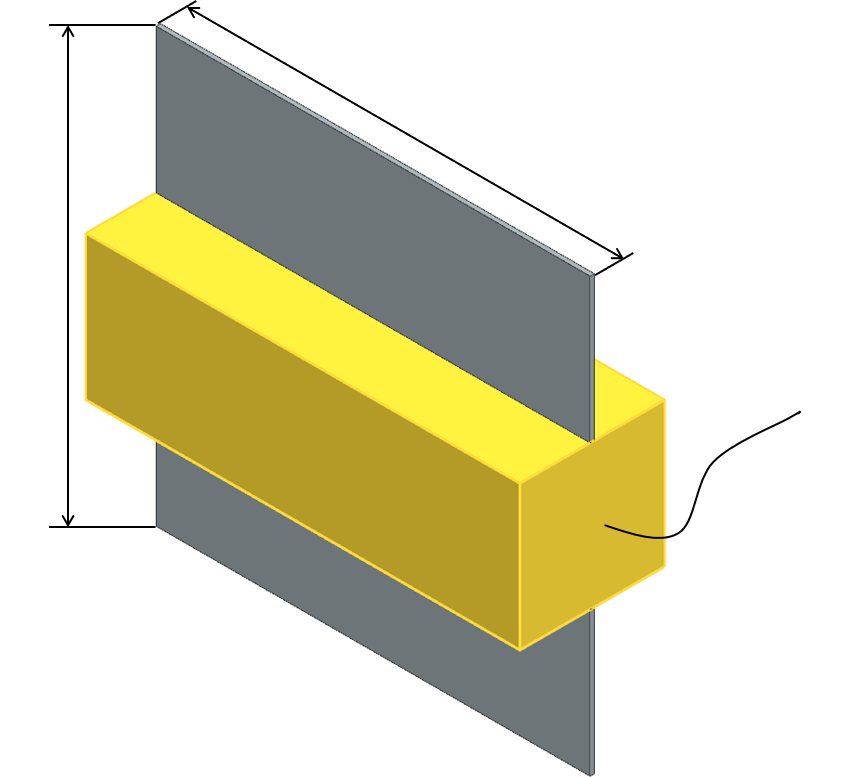}
		\put (90,45){$V_{MB}$}
		\put (2,53){\rotatebox{90}{$h_{max}$}}
		\put (44,80){\rotatebox{-30}{$l_{max}$}}
	\end{overpic}
	\caption{Reference satellite: 3UCubeSat with panels.}
	\label{fig:default_satellite}
\end{figure}
All simulation results are related to a reference satellite under the same conditions in order to evaluate different satellite designs. Due to its universal applicability for various payloads and the size suitable for satellite formations, a 3UCubeSat with one panel each on the top and bottom is used as a reference satellite (see Fig.~\ref{fig:default_satellite}). For compliance with the commonly utilized CubeSats \cite{cubesat}, the mass of the satellite is set to be 5~\si{\kilo\gram}. The reference satellite has a maximum length $l_{max}$ and height $h_{max}$ of 0.3~m each as well as a main body volume $V_{MB}$ of 0.003~\si{\cubic\metre} with main body height and width of both 0.1~m. The width of the panel is equal to 0.003~m.  

The dependence of the atmospheric forces on the surface properties of the satellite is taken into account by considering different energy accommodation coefficients~$\alpha_T$ and corresponding gas-surface-interaction models (GSIM). An overview of the applied surface properties is given in Tab.~\ref{tab:surface_properties}. The chosen gradation steps $\Delta_1$ and $\Delta_2$ are derived from the traditional surface materials used for satellites. $\alpha_{T,1}=1.00$ represents the worst case scenario for a deliberate deflection of the particle. The second value, $\alpha_{T,2}=0.91$ represents a value for the traditional surface materials \cite{Moe,Pilinski} and $\alpha_{T,3}=0.70$ represents improved material properties for diffusely re-emitting materials \cite{pardini}. Therefore, $\Delta_1=0.09$ and $\Delta_2=0.21$, which is equally applied on the energy accommodation coefficients for the consideration of specular reflection as shown in Tab.~\ref{tab:surface_properties}. Additionally, this choice of $\alpha_T$ allows comparison to other literature with similar gradation steps \cite{traub2020influence}. 

\begin{table}[tb]
	\centering
		\begin{tabular}{rl} 
		Diffuse Re-emission & $\alpha_{T,1} = 1.00 \overset{-\Delta_1}{\longrightarrow} 0.91 \overset{-\Delta_2}{\longrightarrow} 0.70$ \\ 
		  Specular Reflection & $\alpha_{T,4} = 0.00 \overset{+\Delta_1}{\longrightarrow} 0.09 \overset{+\Delta_2}{\longrightarrow} 0.30$\\
	\end{tabular}
	\caption{Gradation of the surface properties.}
	\label{tab:surface_properties}%
\end{table}
\subsubsection{Geometry constraints}
For comparability of the results within this work, all design variations have to comply with the following constraints regarding the geometry:
\begin{itemize}
\item \textbf{Constant volume:} Main body volume remains constant.
\item \textbf{Height and length limitations:} Maximum height and length of the reference satellite must not be exceeded.
\end{itemize}
Due to these geometric constraints, the mass of the satellite can be considered constant and is equal to 5~\si{\kilo\gram} according to the reference satellite. Thus, the only variables in Eq.~\ref{eq:ad} and Eq.~\ref{eq:al} are the aerodynamic coefficients and the reference area, and changes in the forces can be directly attributed to the design changes.

\begin{figure*}
	\centering
	\begin{overpic}[width=0.9\textwidth]{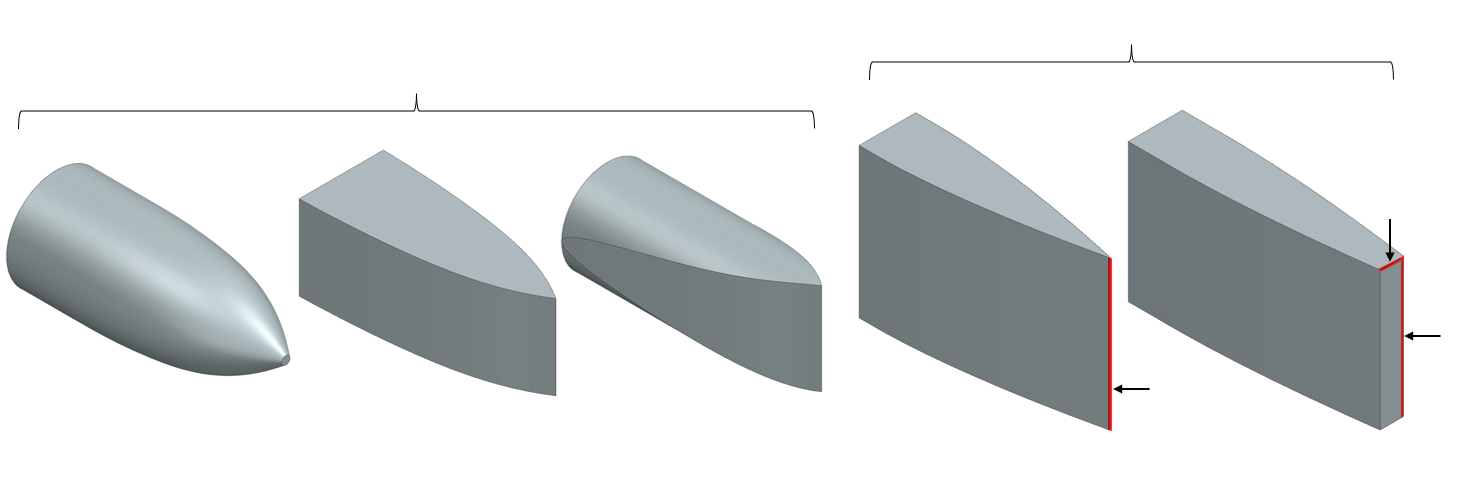}
		\put (20,28){\scriptsize Original Volume Derivations}
		\put (4,5){\scriptsize Geometry A}
		\put (24,5){\scriptsize Geometry B}
		\put (42,5){\scriptsize Geometry C}
		\put (68,31){\scriptsize Extended Volume Derivations}
		\put (63.5,2){\scriptsize B+ $h_{MB}$}
		\put (82,2){\scriptsize B+ $w_{N}$, $h_{MB}$}
		\put (79.3,6.3){\footnotesize $h_{MB}$}
		\put (99,10){\footnotesize $h_{MB}$}
		\put (94,19){\footnotesize $w_{N}$}
	\end{overpic}
	\caption{Possible 3D volume derivations for the 2D optimization.}
	\label{fig:Opt_Tool_Geos}
\end{figure*}

\begin{figure*}
	\centering
	\begin{overpic}[width=0.8\textwidth]{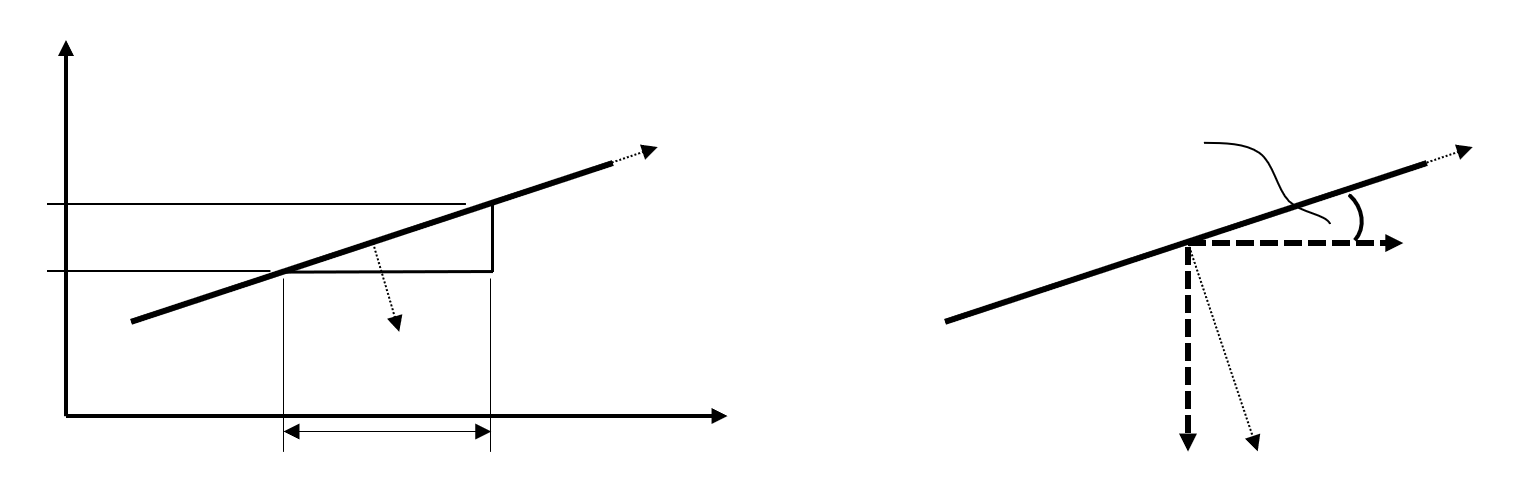}
		\put (3.5,30){y}
		\put (48,3.75){x}
		\put (43,22){$x_{local}$}
		\put (22,8){$y_{local}$}
		\put (-6.5,13){$f(i-1)$}
		\put (-2.5,17.5){$f(i)$}
		\put (23.5,0.5){d$x$}
		\put (92,15){$\rightarrow_{drag}$}
		\put (76.5,-1){$\downarrow_{lift}$}
		\put (74.5,21.5){$\Theta_i$}
		\put (83,2){$y_{local}$}
		\put (96,22){$x_{local}$}
	\end{overpic}
	\caption{Definition of the 2D optimization tool function values according to Hild \cite{Hild.2021}.}
	\label{fig:Opt_Tool}
\end{figure*}
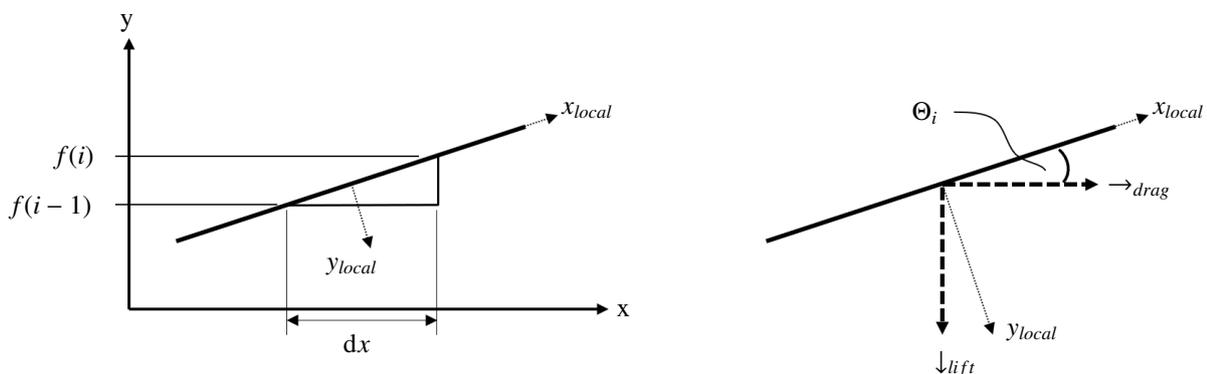
\subsection{Utilized tools}
The utilized open-source tools as well as an extension for a Matlab 2D profile optimization tool for the derivation of satellite geometries are introduced in the following sections.
\subsubsection{Determining the atmospheric forces}
Since numerous investigations of the influences of different design parameters are necessary, the calculation time for the aerodynamic forces per satellite model plays an important role. Therefore, the Matlab toolkit "ADBSat", which was developed at the University of Manchester \cite{Sinpetru.2022}, was used to study the effects of different designs on the aerodynamic forces. It represents a means of calculating the acting forces with little computational effort. However, this method is not suitable for concave surfaces, since multiple reflections of the incoming particles cannot
be taken into account. For selected and promising geometries as well as geometries causing multiple reflections, their actual performance was determined using the more computationally intensive but proven Direct Simulation Monte Carlo (DSMC) method, which is available within the gas-kinetic simulation framework "PICLas", developed at the University of Stuttgart \cite{Fasoulas.2019}. For the PICLas simulations, the necessary variable hard sphere parameters for the species listed in Tab.~\ref{tab:Atmosphere_Data} were taken from Bird \cite{Bird.1994}. A comparison of ADBSat and PICLas results can be found in \ref{sec:PIClasADBSat}.

\subsubsection{2D profile optimization}
\label{sec:MatlabTool}
For optimizing and obtaining the satellite models, an already existing Matlab 2D optimization tool for reduced drag \cite{Hild.2022} was utilized and adapted in the scope of this work. This tool was developed for an analytical determination of a 2D profile with minimum drag for a given length and volume. A preceding determination of the type of volume derivation of the profile is necessary in order to comply with the given volume restrictions during the profile optimization. In the given optimization tool, three different volume derivation options are possible as shown on the left in Fig.~\ref{fig:Opt_Tool_Geos}. Whereas geometry option~A is obtained by rotating the profile around the x-axis (longitudinal axis), geometry option~B is obtained by mirroring the profile at the x-axis and then extruding symmetrically with the value of the largest cross-section width. Thereby, geometry option~B has a squared rear side by default. Geometry option~C is obtained by intersecting the extruded profile of option B with a cylinder with the given length and a radius equal to the largest cross-section width of the optimized profile. Additionally, the application of a tail geometry is possible, in which the height of the 2D profile is reduced again from a certain longitudinal position. Each individual profile optimization complies with the given volume restriction of $V_{MB}=0.003$~\si{\cubic\metre}. Within this work, the 2D profile optimization tool was extended as presented in the following:

\textbf{Additional geometry options:}   
The extrude option B was modified to consider the following two additional user inputs (see Fig.~\ref{fig:Opt_Tool_Geos} right):
\begin{enumerate}
    \item \textbf{Extrusion height $\bm{h_{mb}}$:} For the application of the optimization tool on a given structure such as the panels of this work's reference satellite, the height of the extrusion can be set as an input parameter.
    \item \textbf{Nose width $\bm{w_N}$:} Whereas it might be aerodynamically advantageous to have a sharp or pointy nose geometry, it could pose a problem for accommodating the payload and manufacturing of the satellite. Hence a definition of a nose width during the optimization process was implemented.
\end{enumerate} 
\textbf{Lift optimization:} Within the 2D optimization tool, a derivation of Sentman's equation was used \cite{Hild.2022}. Since the Matlab tool is aiming on optimizing a 2D profile, it followed for Sentman's equation that the two direction cosine $\eta, t = 0$. Considering the definition of the area element on the profile as shown in Fig.~\ref{fig:Opt_Tool}, the direction cosine between the local x and y axis and the direction of the drag force is set as follows in the original tool:
\begin{align}
	k_d & = \cos(\Theta_{i}) \\
	l_d & = \sin(\Theta_{i})
\end{align}
with
\begin{equation}
	\Theta_i = \text{atan}\left(\frac{f(i)-f(i-1)}{\text{d}x}\right).
\end{equation}
By setting the two direction cosine according to the desired force direction of the lift force (see Fig.~\ref{fig:Opt_Tool}) to
\begin{align}
k_L & = \cos\left(\Theta_{i} + \frac{\pi}{2}\right) \text{ and}\\
l_L & = \sin\left(\Theta_{i} + \frac{\pi}{2}\right),
\end{align}
an optimization of a given profile with regard to increasing the lift force experienced by the profile is now possible as well. It should be noted, that therefore the sign of the original optimization function has to be inverted, since originally a decrease in the acting force was desired.

\begin{figure*}
  \centering
    \begin{subfigure}{0.45\textwidth}
    \centering
      \includegraphics{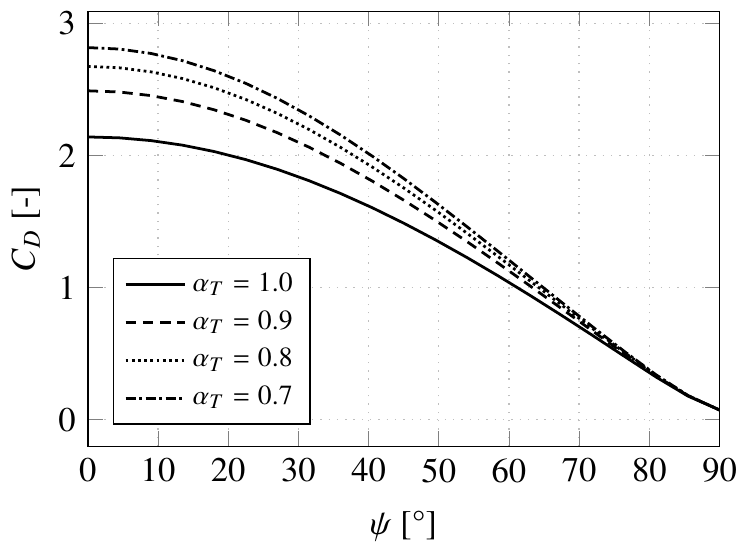}
    \end{subfigure}
    \begin{subfigure}{0.45\textwidth}
    \centering
      \includegraphics{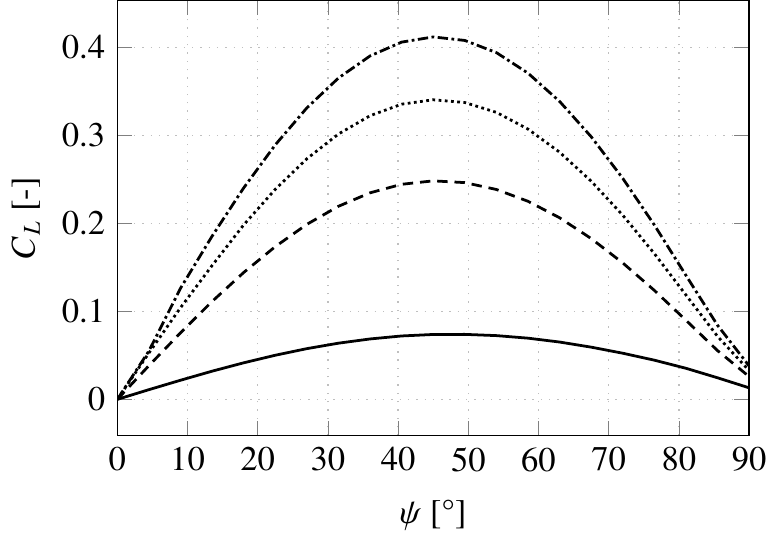}
    \end{subfigure}
\caption{Aerodynamic coefficients for a flat plate depending on the angle $\psi$ between flow and surface normal according to Sentman's model.}
\label{fig:coeff}
\end{figure*}

\begin{figure*}
\centering
  \subfloat[Raster 1\label{subfig:Design1a}]{%
    \resizebox{0.31\textwidth}{!}{\includegraphics[width=0.5\textwidth]{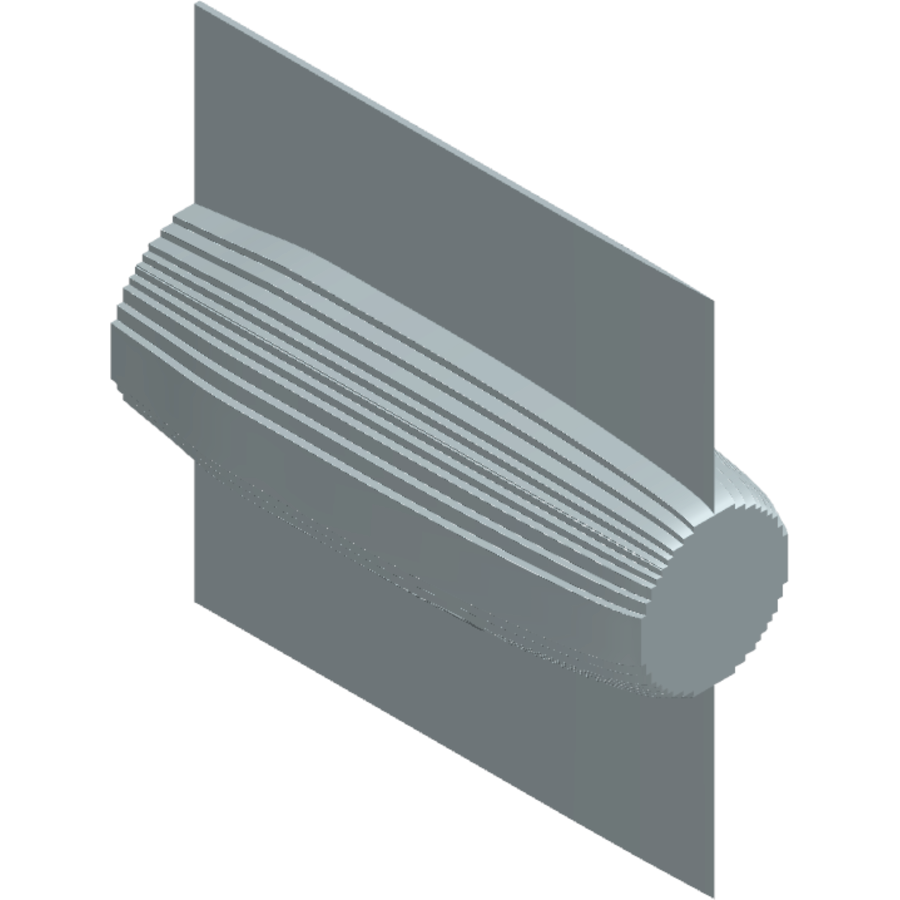}}%
  } \quad
  \subfloat[Raster 2\label{subfig:Design1b}]{%
    \resizebox{0.31\textwidth}{!}{\includegraphics[width=0.5\textwidth]{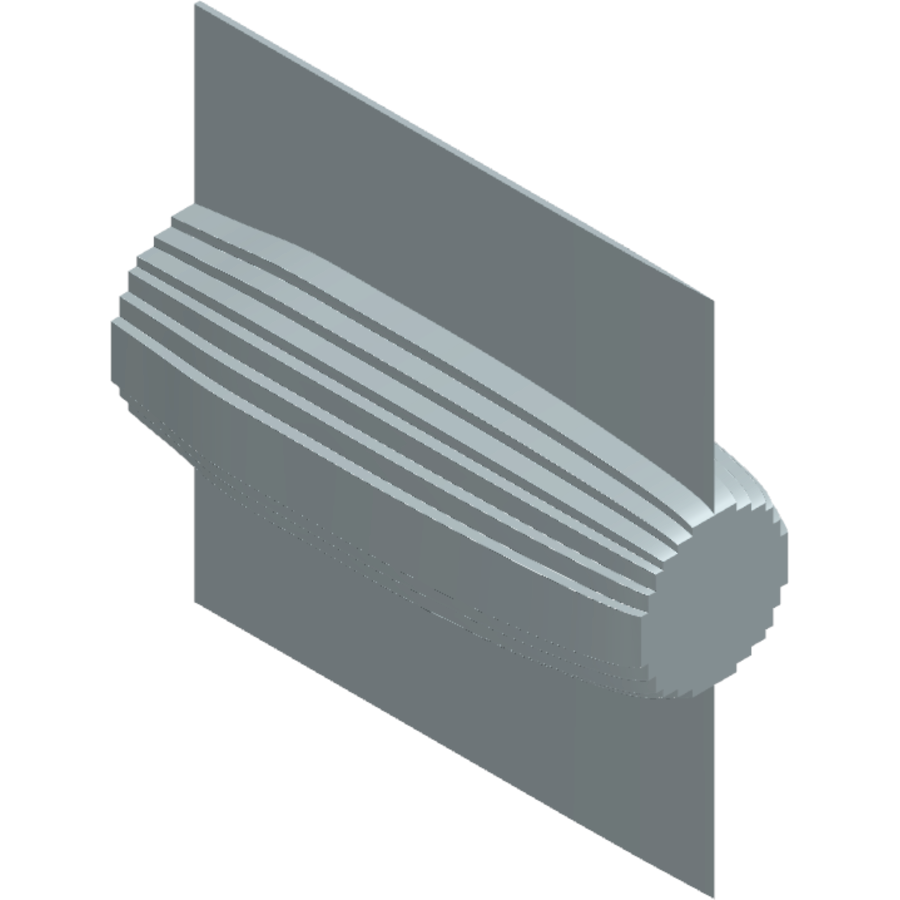}}%
  } \quad
  \subfloat[Raster 3\label{subfig:Design1c}]{%
    \resizebox{0.31\textwidth}{!}{\includegraphics[width=0.5\textwidth]{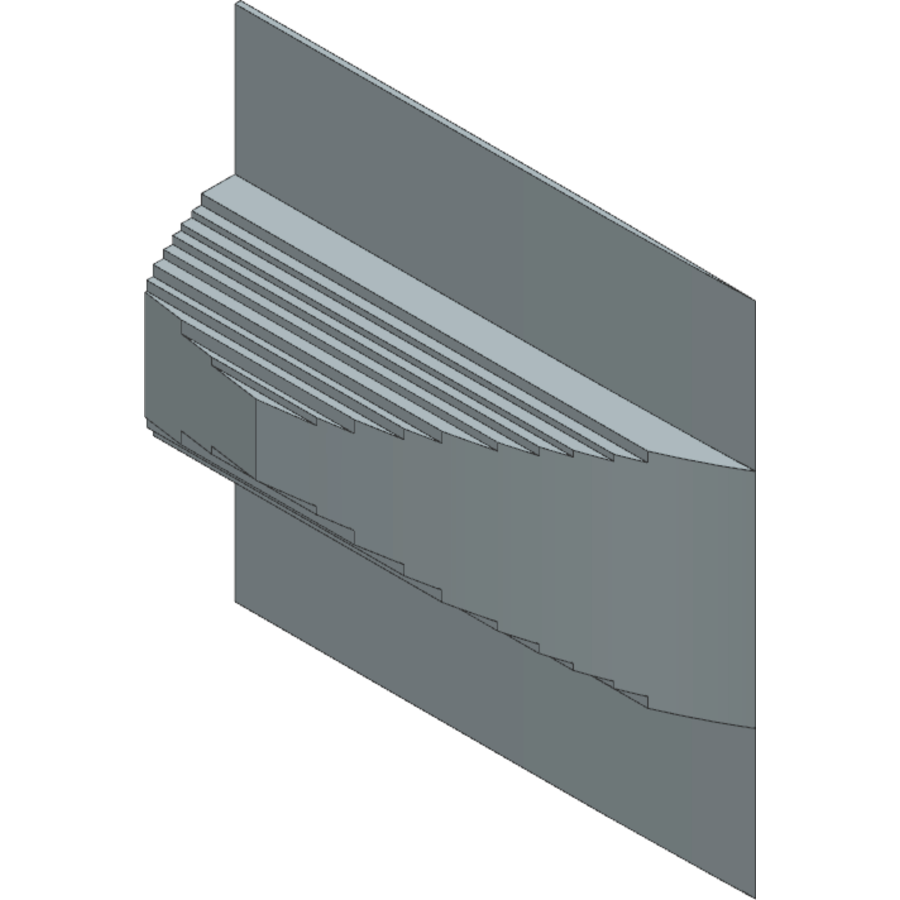}}%
  }
  \\
  \subfloat[MultRef 1\label{subfig:Design2a}]{%
    \resizebox{0.31\textwidth}{!}{\includegraphics[width=0.5\textwidth]{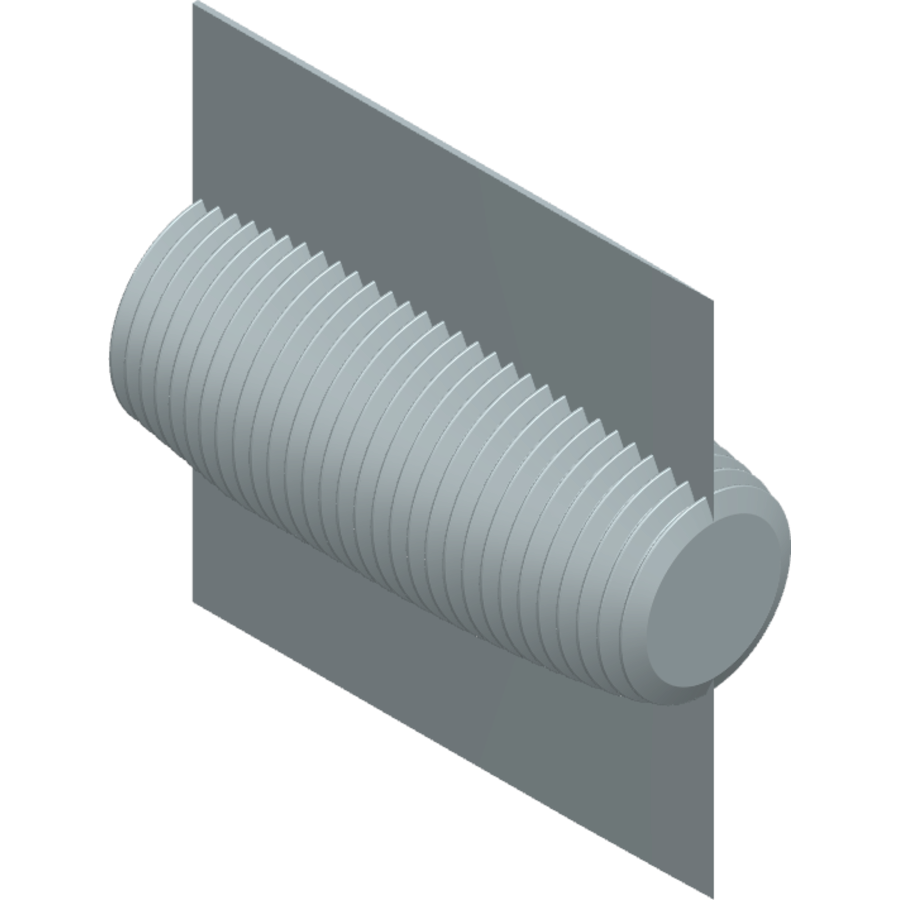}}%
  } \quad
  \subfloat[MultRef 2\label{subfig:Design2b}]{%
    \resizebox{0.31\textwidth}{!}{\includegraphics[width=0.5\textwidth]{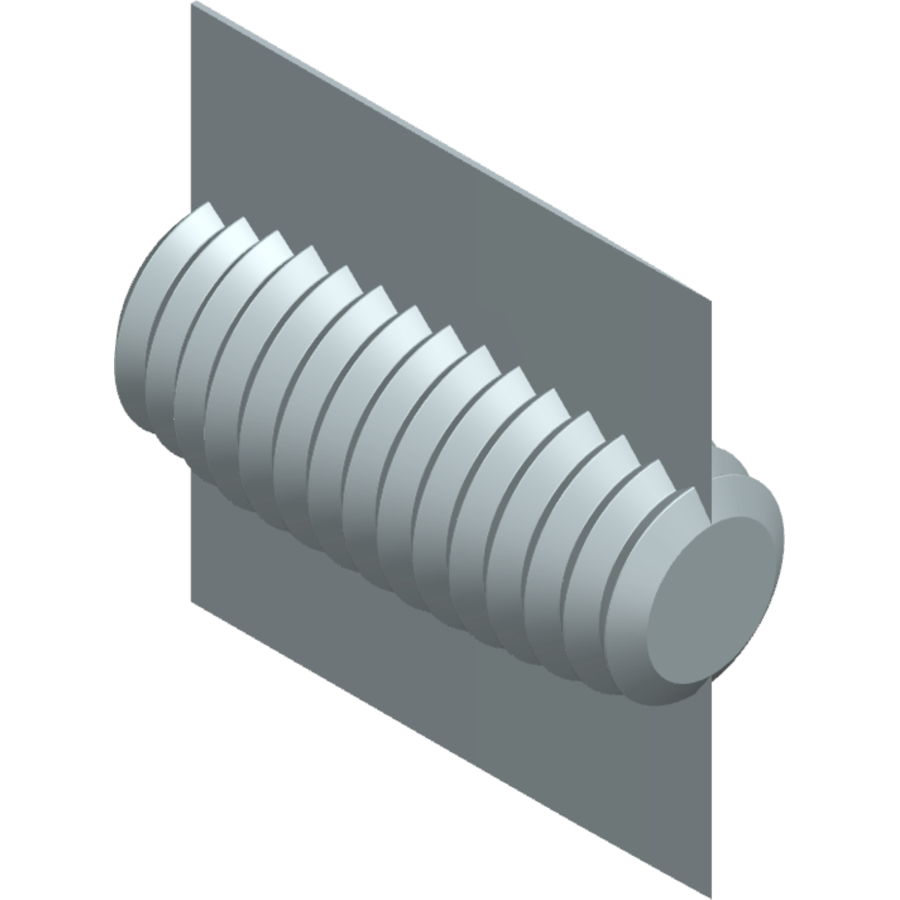}}%
  }\quad  
   \subfloat[Lift 1\label{subfig:Design3a}]{%
    \resizebox{0.31\textwidth}{!}{\includegraphics[width=0.5\textwidth]{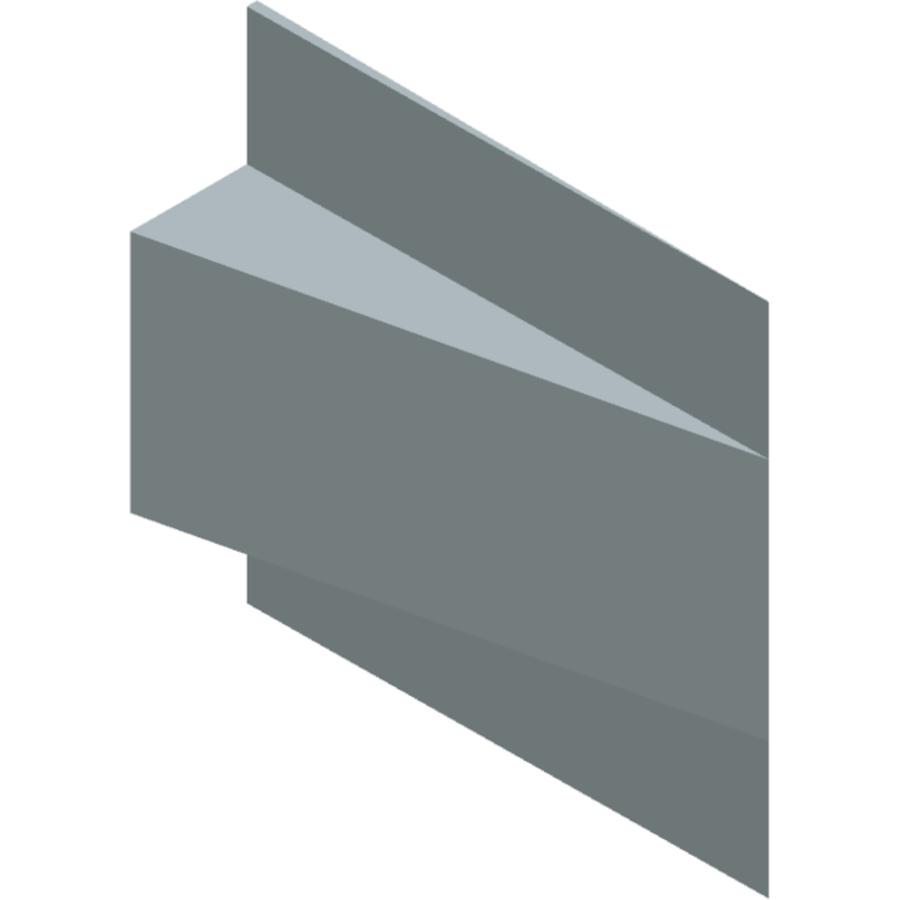}}%
      }
  \\
  \subfloat[Lift 2\label{subfig:Design3b}]{%
    \resizebox{0.31\textwidth}{!}{\includegraphics[width=0.5\textwidth]{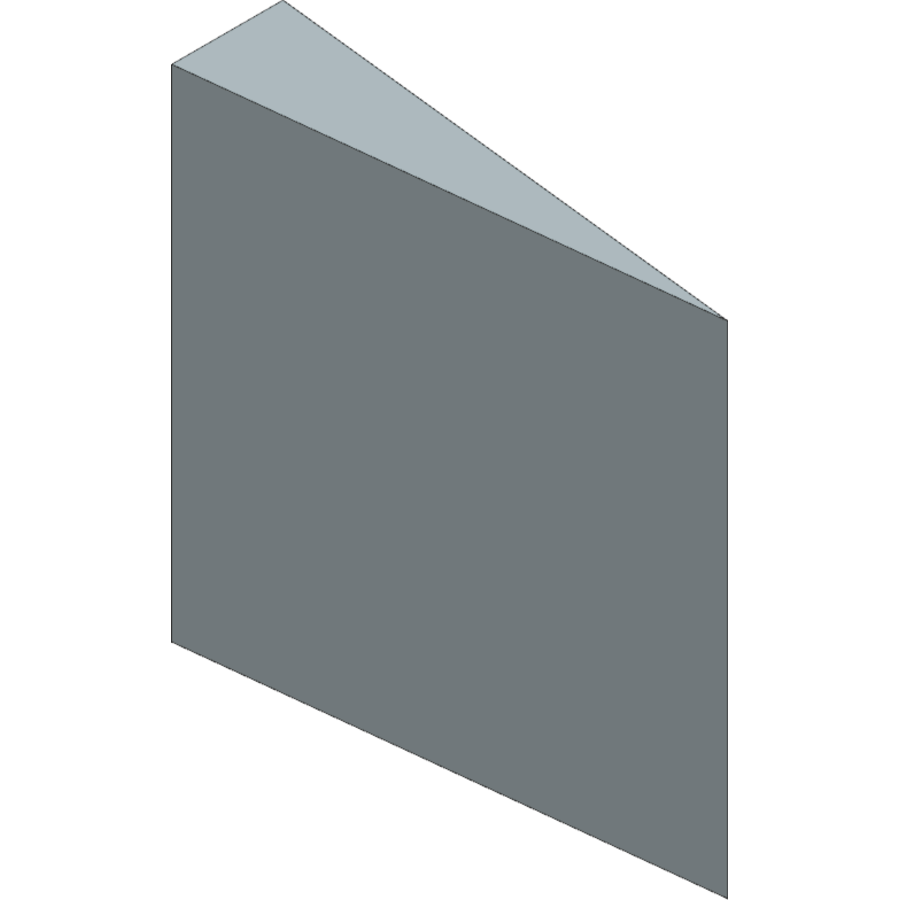}}%
    } \quad
    \subfloat[SpecOpt\label{subfig:Design4a}]{%
    \resizebox{0.31\textwidth}{!}{\includegraphics[width=0.5\textwidth]{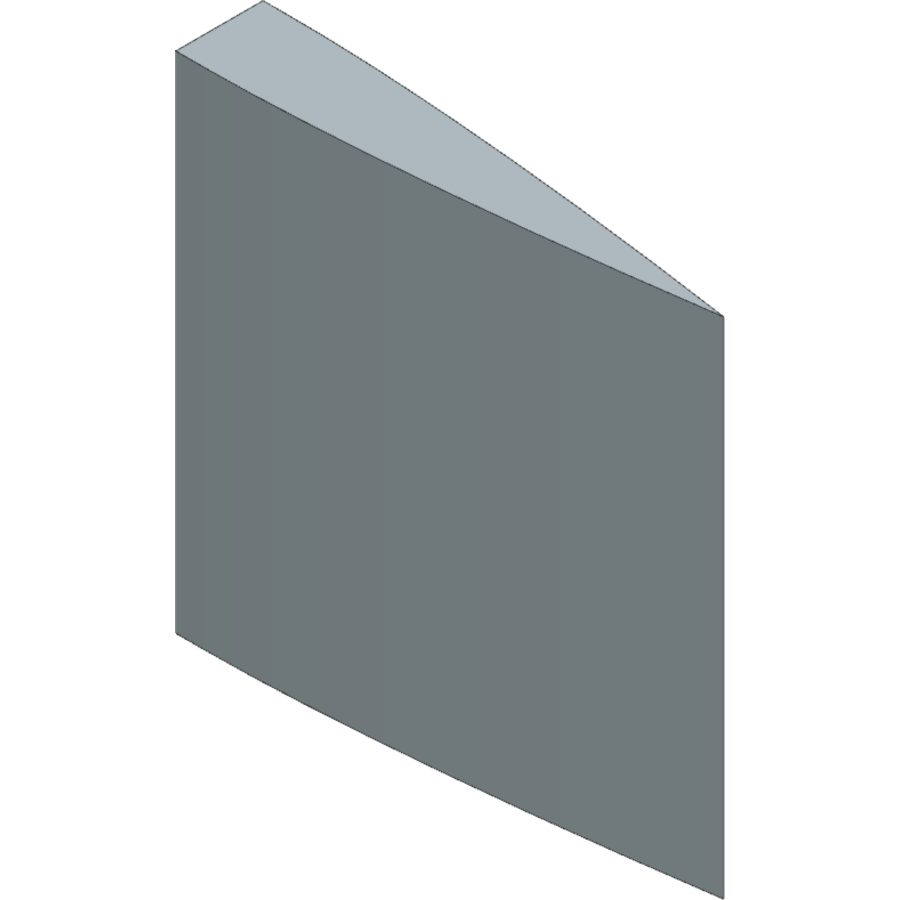}}%
  } \quad
  \subfloat[SpecPrac\label{subfig:Design4b}]{%
    \resizebox{0.31\textwidth}{!}{\includegraphics[width=0.5\textwidth]{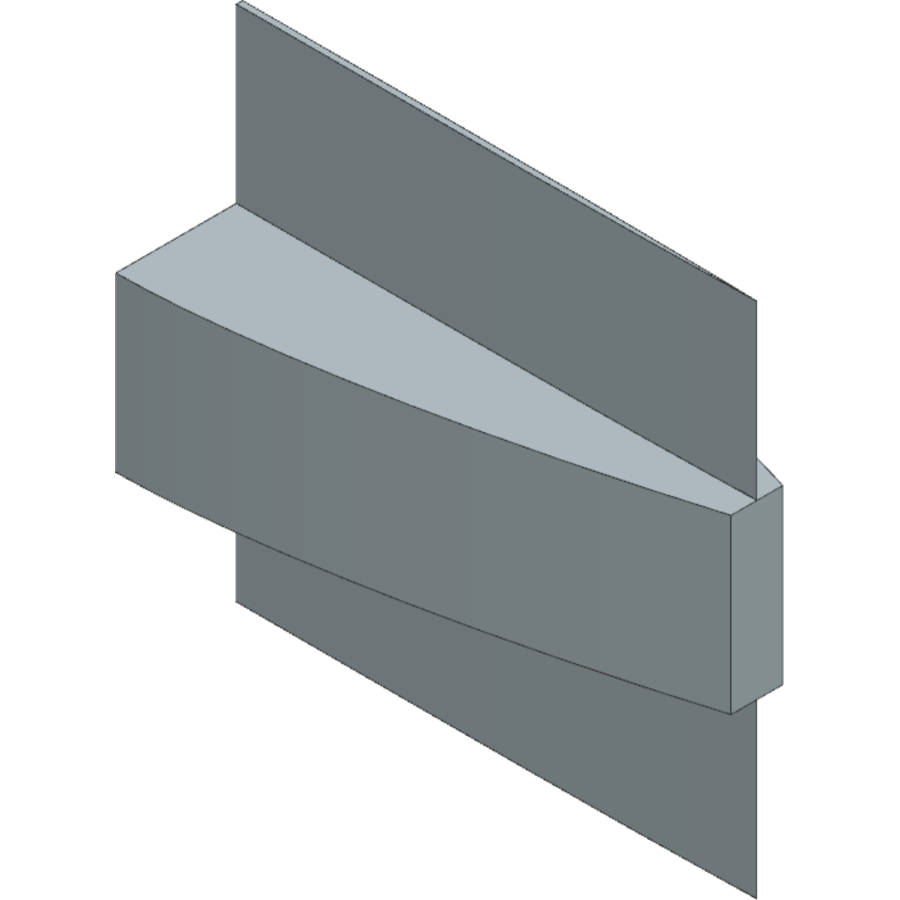}}%
    } 
\caption{Selection of designs evaluated with PICLas and ADBSat.}
\label{fig:overview}
\end{figure*}
\subsection{Optimization steps}
The optimization process presented in this subsection is equally executed for the assumption of diffuse re-emission and specular reflection. In a first step, the design is varied with the aim of optimizing the design for solely differential drag or solely differential lift, in order to investigate the respective important design characteristics. In a second step, the design characteristics are combined for achieving a best possible trade-off according to the set priorities. 

The optimization process for deriving a design optimal for differential drag is divided into the following steps:
\begin{description}
\item[Drag 1a - Decreasing minimum drag (main body):] The main body is optimized for minimum drag in the nominal flight configuration utilizing the extended Matlab 2D profile optimization tool (EPOT) while maintaining the maximum length of 30~cm and the main body volume of 0.003~\si{\cubic\metre}.
\item[Drag 1b - Decreasing minimum drag (panel):] The panel is optimized for minimal drag in the nominal flight configuration using the EPOT, considering a given extrusion height of 30~cm. The trivial solution, omitting the panel, is not considered within this work as it would greatly reduce the achievable control forces. Thus, the size of the panel remains $l_{max} \cdot h_{max}$.
\item[Drag 2 - Increasing maximum drag:] The loss in maximum drag due to the first and second optimization steps has to be kept as small as possible. Considering the optimum angle for the maximum of $C_D$ (see Fig.~\ref{fig:coeff}) and the set geometry restrictions, the following approaches can be derived:
\begin{itemize}
\item \textbf{Increasing number of area elements with $\bm{C_{D,max}}$:} For an AOS of 90°, $A_{ref}$ cannot be further increased due to the given $l_{max}$ and $h_{max}$ and thus, the overall $C_D$ of the whole body needs to be maximized. A possible solution for increasing $C_D$ is implementing the largest perpendicular area to the flow in the maximum drag configuration. Within this work, this was achieved by changing the panel position and approximating the satellite shape with vertical and horizontal area elements (see Figs.~\ref{subfig:Design1a}-\ref{subfig:Design1c}).
\item \textbf{Multiple reflections:} For a consideration of the whole geometry and not only single area elements (as within ADBSat), a geometry promoting multiple reflections poses another possible approach to increase the drag. This approach was implemented in the design by overlaying the optimized 2D profile with a zigzag curve for particle-trapping concave areas (see Figs.~\ref{subfig:Design2a}-\ref{subfig:Design2b}).
\end{itemize}
\end{description}
Opposite to the design optimization process for differential drag, where a high difference of $\vec{f}_{D,min}$ and $\vec{f}_{D,max}$ but generally lower absolute values of $\vec{f}_{D,min}$ and $\vec{f}_{D,max}$ are desired, the goal for optimizing the design for differential lift only consists of increasing $\vec{f}_{L,max}$. The process for optimizing the design for differential lift was divided into the two steps:
\begin{description}
\item[Lift 1a - Increasing maximum lift:] For achieving higher lift forces, the profile obtained using the EPOT lift optimization function (see Fig.~\ref{subfig:Design3a} and \ref{subfig:Design3b}) was implemented in the design. 
\item[Lift 1b - Eliminating lift reducing areas:] When considering the reference satellite with its frontal area it is apparent, that for an AOS $\approx$ 45° (i.e. maximum $C_L$ value for the panels, see Fig.~\ref{fig:coeff}), the frontal area and the velocity vector form an angle $\approx -$45°. Therefore, the frontal area experiences lift as well, but in the opposite direction as desired. Hence, when dealing with different profiles than the optimized one, areas experiencing lift in the opposite direction than desired must be decreased.
\end{description}

Each design was generally tested in a variety of AOA and AOS in order to determine the orientation achieving the highest drag and lift force, referred to as the maximum drag and maximum lift configuration, respectively. The same was applied to the reference satellite. Therefore, in order to compare the optimization in drag and lift forces, the respective configurations with the maximum achievable values are compared to each other.
\section{Design optimization - diffuse re-emission}
\label{sec:process1}
The results of a design variation are generally related to the reference satellite with the same surface properties, if not stated otherwise. The gas-surface-interaction model applied for the ADBSat calculations within this section is Sentman's model \cite{Sentman.1961}. 
\subsection{Differential drag optimization} 
Within the design considerations for the use of differential drag, the influence of general shape and tail geometry, surface structures and panel position has been analyzed, and the results are presented in the following sections.
\subsubsection{General shape and tail geometry}
\begin{figure*}
\centering
\begin{subfigure}{.36\textwidth}
  \centering
  \includegraphics{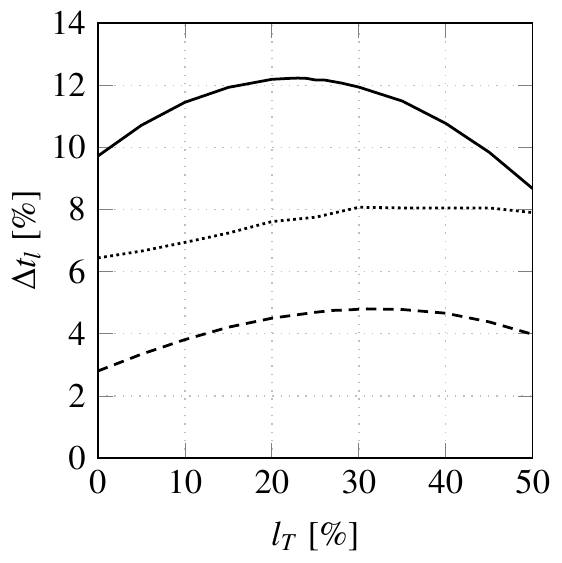}
  \caption{$\alpha_T=1.00$}
  \label{subfig:diffuse_tails_100}
\end{subfigure}
\begin{subfigure}{.30\textwidth}
  \centering
  \includegraphics{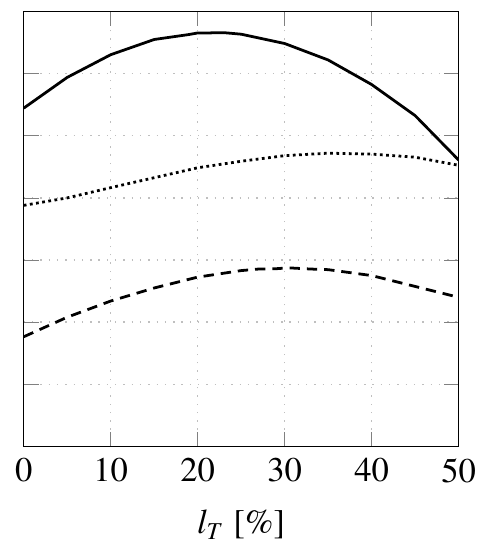}
  \caption{$\alpha_T=0.91$}
  \label{subfig:diffuse_tails_091}
\end{subfigure}
\begin{subfigure}{.30\textwidth}
  \centering
  \includegraphics{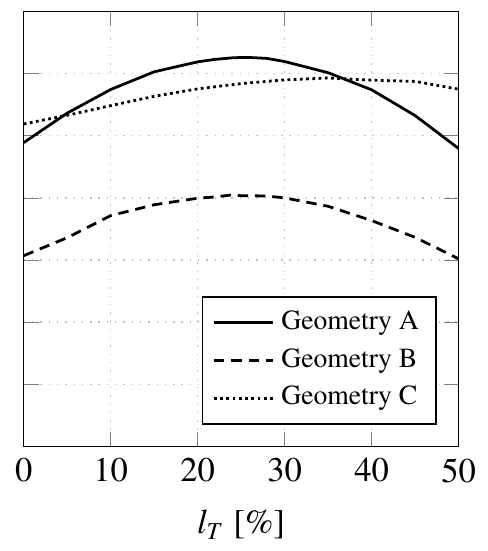}
  \caption{$\alpha_T=0.70$}
  \label{subfig:diffuse_tails_070}
\end{subfigure}
\caption{Dependence of the lifetime on general geometry and tail length.}
\label{fig:diffuse_tails}
\end{figure*}

\begin{figure*}
\centering
\begin{subfigure}{.36\textwidth}
  \centering
  \includegraphics{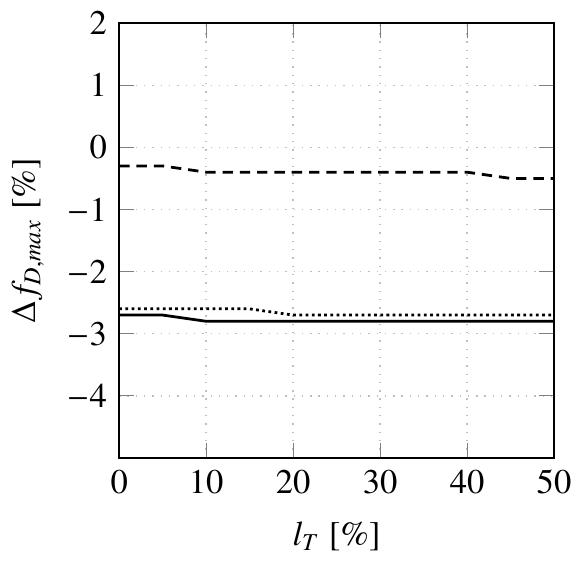}
  \caption{$\alpha_T=1.00$}
  \label{subfig:diffuse_tails_drag_100}
\end{subfigure}
\begin{subfigure}{.30\textwidth}
  \centering
  \includegraphics{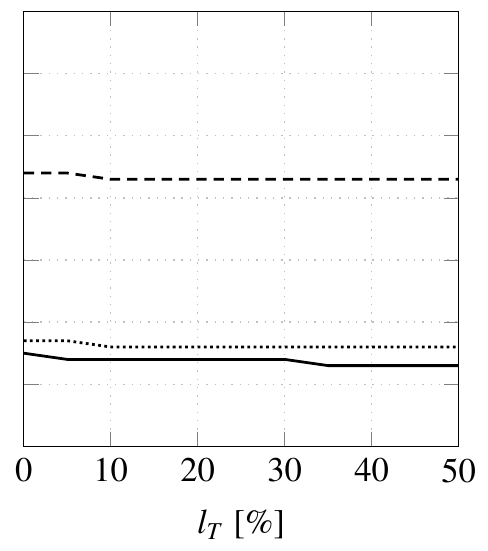}
  \caption{$\alpha_T=0.91$}
  \label{subfig:diffuse_tails_drag_091}
\end{subfigure}
\begin{subfigure}{.30\textwidth}
  \centering
  \includegraphics{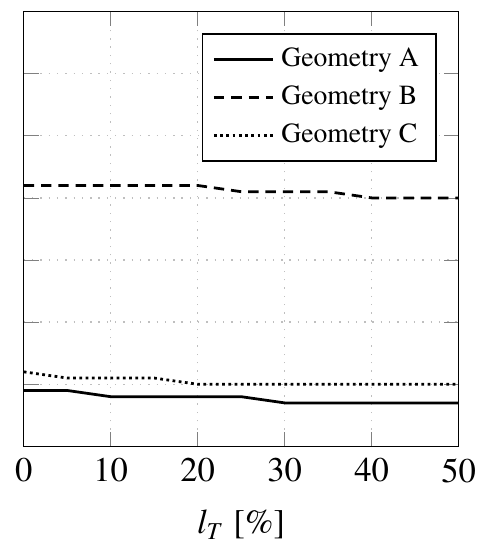}
  \caption{$\alpha_T=0.70$}
  \label{subfig:diffuse_tails_drag_070}
\end{subfigure}
\caption{Dependence of maximum drag on general geometry and tail length.}
\label{fig:diffuse_tails_drag}
\end{figure*}

\begin{figure*}
\centering
\begin{subfigure}{.36\textwidth}
  \centering
    \includegraphics{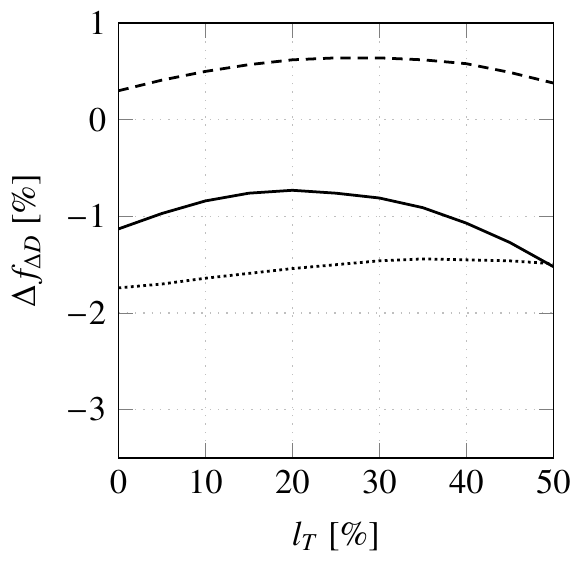}
  \caption{$\alpha_T=1.00$}
  \label{subfig:diffuse_tails_diffdrag_100}
\end{subfigure}
\begin{subfigure}{.30\textwidth}
  \centering
    \includegraphics{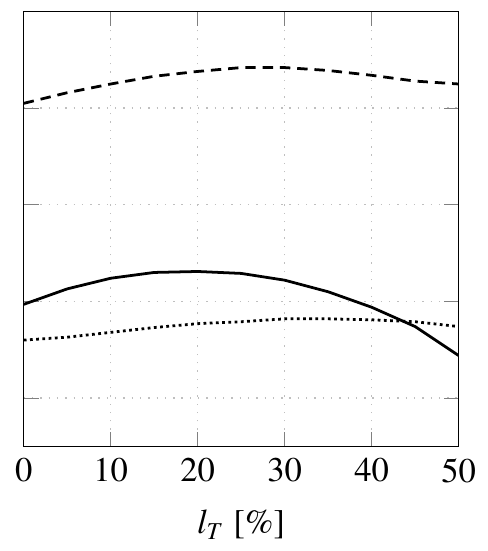}
  \caption{$\alpha_T=0.91$}
  \label{subfig:diffuse_tails_diffdrag_091}
\end{subfigure}
\begin{subfigure}{.30\textwidth}
  \centering
    \includegraphics{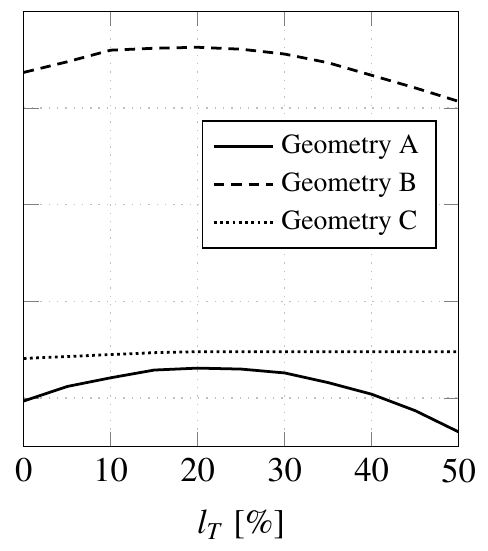}
  \caption{$\alpha_T=0.70$}
  \label{subfig:diffuse_tails_diffdrag_070}
\end{subfigure}
\caption{Dependence of differential drag on general geometry and tail length.}
\label{fig:diffuse_tails_diffdrag}
\end{figure*}

As investigated by Hild \cite{Hild.2021} and Walsh \cite{Walsh.}, it can be advantageous for the overall lifetime of the satellite to apply a tail geometry on slender bodies. In the following, the effect of a tail geometry for the three different volume derivation options A, B and C is investigated in order to find the optimum combination for an increased lifetime. At the same time, their influence on the achievable maximum drag and thereby the overall differential drag is assessed. The influence of the tail length on the lifetime in nominal flight configuration and experienced drag in maximum drag configuration as well as the resulting differential drag for different $\alpha_T$ can be seen in Fig.~\ref{fig:diffuse_tails}, Fig.~\ref{fig:diffuse_tails_drag} and Fig.~\ref{fig:diffuse_tails_diffdrag}. The data shown was obtained using ADBSat.


It is apparent, that the optimum tail length generally depends on the volume derivation option and the surface properties. Whereas the geometry option A is promising for high $\alpha_T$, the increase in lifetime for geometry options B and C generally increases with decreasing $\alpha_T$. The three following geometry options and tail lengths are optimal concerning the lifetime for the given cases of surface properties: for $\alpha_T=1.00$ and $\alpha_T=0.91$ geometry A with 23~\% tail length and for $\alpha_T=0.70$ geometry A with 25~\% tail length. However, as shown in Fig.~\ref{fig:diffuse_tails_drag}, geometry option B experiences the higher drag forces in maximum drag configuration as expected due to its vertical side surfaces, which are perpendicular to the flow in maximum drag configuration and thus correspond to the highest $C_D$~values. Even though geometry option~A can reach a lifetime increase of up to 13\%, the negative effect on the maximum achievable drag by the rounded side surfaces leads to an overall loss in differential drag compared to the reference satellite as visible in Fig.~\ref{fig:diffuse_tails_diffdrag}. The increase in lifetime with geometry option B is smaller than for A and C, but due to the generally higher maximum drag, an increase in differential drag compared to the reference satellite is possible. However, as the priority of this work is to increase the lifetime, options for increasing the maximum drag while maintaining the geometry characteristics of geometry option A were investigated.
\subsubsection{Surface structures}
Since the main body optimization with the EPOT led to side areas with different orientations to the flow and therefore smaller $C_D$ values, the maximum drag experienced is smaller compared to the reference satellite with completely vertical side surfaces. In order to increase the $C_D$ values of the side surfaces in the maximum drag configuration while maintaining the overall optimized profile shape obtained by the EPOT, a rasterized cross section was added to the optimized profile in order to approximate the curved surface by horizontal and vertical surface elements. Therefore, the frontal circular area, which was swept along the optimized profile, is rasterized using the midpoint circle algorithm\footnote{The midpoint circle algorithm is an algorithm used for determining the points of a rasterized circle.} to two different raster step sizes as shown in Fig.~\ref{subfig:Design1a} and Fig.~\ref{subfig:Design1b}. Here, Raster~1 has a step size of $r_{front}/14.3$ , with $r_{front}$ being the radius of the frontal area, which is a result of the EPOT for given volume restriction, $l_{max}$, $\alpha_T$ and geometry option. With this step size, the geometry constraint of constant main body volume remains met. Raster~2 has a step size of $r_{front}/10$. Another possible means to increase the maximum drag is to use surface structures promoting multiple reflections. Therefore, the geometries shown in Fig.~\ref{subfig:Design2a} and Fig.~\ref{subfig:Design2b}, which are derived by overlaying the optimized profile for geometry option A with a zigzag-curve with an extrema distance~$d_{ex}$ of 5~mm (design MultRef~1) and 10~mm (design MultRef~2) as well as extrema of $\pm5$~mm, were investigated further.

\begin{figure*}
\centering
\begin{subfigure}{.49\textwidth}
    \centering
    \includegraphics{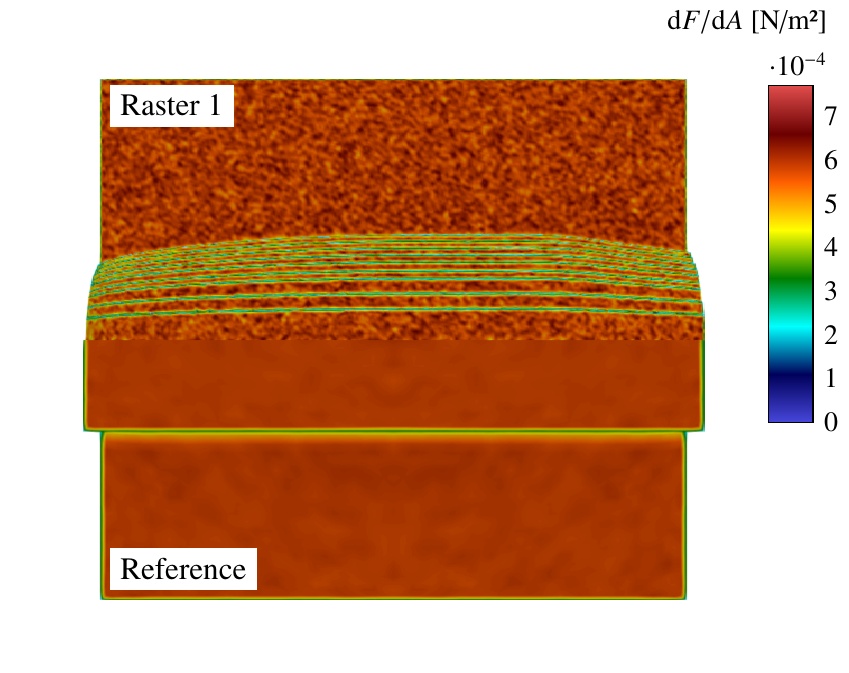}
  \caption{Top: satellite design with additional vertical side surfaces \\ Bottom: reference satellite}
  \label{subfig:piclas_stairs}
\end{subfigure}
\begin{subfigure}{.49\textwidth}
  \centering
    \includegraphics{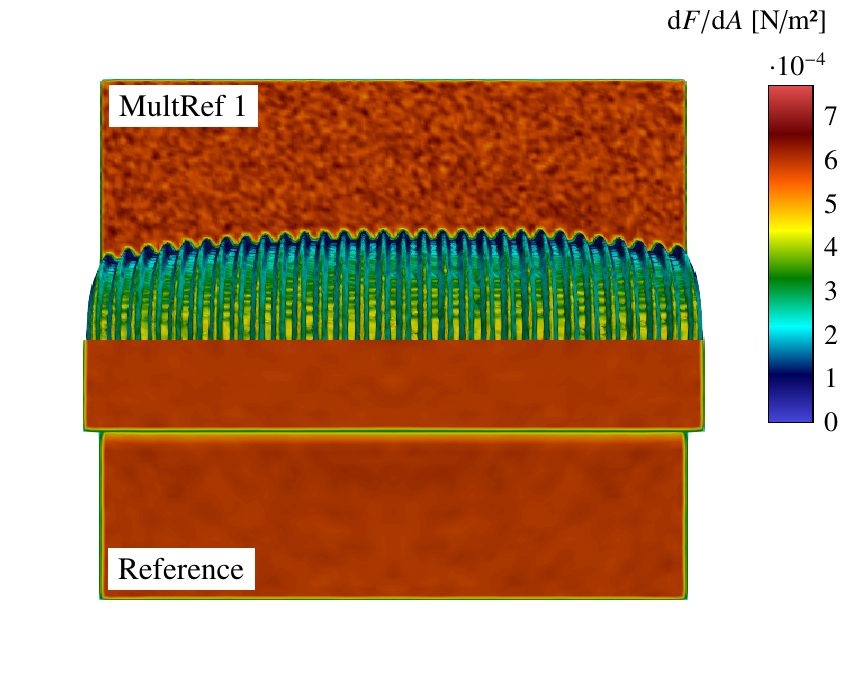}
  \caption{Top: satellite design promoting multiple reflections \\ Bottom: reference satellite}
  \label{subfig:piclas_kres}
\end{subfigure}
\caption{Force distribution shown in head on view in maximum drag configuration for $\alpha_T=1.00$.}
\label{fig:piclas_drag}
\end{figure*}

The results for the above presented satellite designs are listed in Tab.~\ref{tab:drag_stairs_raster}. Here, the lifetime was estimated in the nominal flight configuration and the maximum drag in its respective configuration. The data shown was obtained using PICLas and is referring to the data from the reference satellite obtained using PICLas as well. 
\begin{table}
	\centering
		\begin{tabular}{l|ccc} 
		Design & $\Delta t_l$ & $\Delta f_{D,max}$ & $\Delta f_{\Delta D}$ \\ \hline 
		Raster 1 & +12.26 \% & -0.32 \% & +2.44 \% \\ 
		Raster 2 & +11.81 \% &  -0.31 \% & +2.34 \%\\
		$d_{ex}=5$ mm & +3.27 \% & -1.55 \% & -1.13 \%\\ 
		$d_{ex}=10$ mm & +0.35 \% & -0.50 \% & -0.54 \%\\ 							
	\end{tabular}
	\caption{Data for the tested designs with surface structures, $\alpha_T=1.00$.}
	\label{tab:drag_stairs_raster}%
\end{table}
It can be seen, that the loss in maximum drag is smaller for the design with additional vertical side surfaces than the designs promoting multiple reflections. The force distribution of the different designs compared to the reference satellite is shown in Fig.~\ref{fig:piclas_drag} for $\alpha_T=1.00$ and obtained using PICLas. As the area element's $C_D$ value is the highest for the vertical surfaces, these surfaces experience the highest drag forces. However, the effect of multiple reflections for the tested design is lower than expected and the additional decrease in vertical side surface area leads to an overall loss in experienced drag.
The expected effect of multiple reflections increasing the drag in maximum drag configuration for the tested design did not appear. However, the results for the minimum drag in nominal flight configuration differed up to 25~\% compared to the results obtained using ADBSat. Therefore, the effect of multiple reflections did occur, but not in the desired way. It can be assumed, that the utilization of structures promoting multiple reflections for an increased maximum drag force is generally possible, although the tested design is not suitable in this respect. The rasterized cross section on the other side did not lead to a loss in the generally increased lifetime by the optimized profile, and additionally increased the effective side area and thereby the differential drag. In the case of a chosen geometry option C, the rasterized cross section can be applied as well for increasing the vertical side surfaces (see Fig.~\ref{subfig:Design1c}).
\subsubsection{Panel position}
\begin{figure}
\centering
     \begin{overpic}[width=0.45\textwidth]{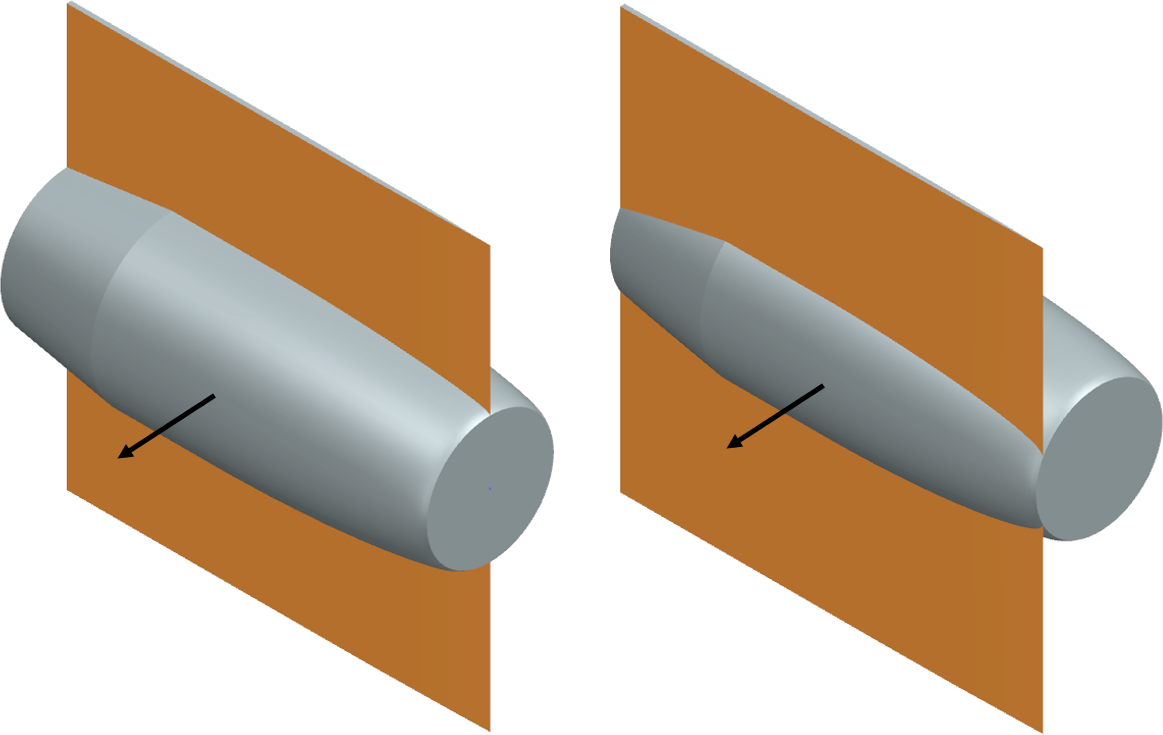}	
		\put (1,19){$\vec{v}_{rel}$}
		\put (58,19){$\vec{v}_{rel}$}		
	\end{overpic}
\caption{Increasing vertical side surfaces by changing the panel position.}
\label{fig:panpo_cad}
\end{figure}

\begin{figure}
  \centering
  \  \includegraphics{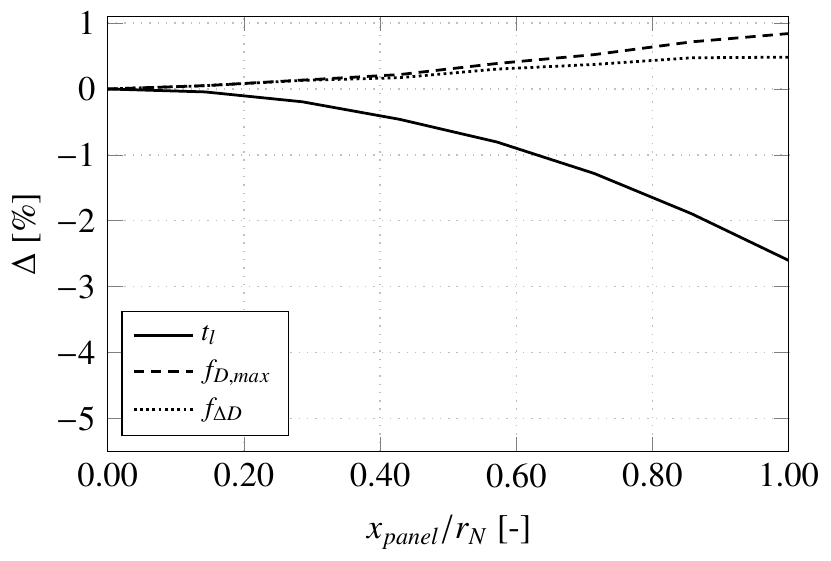}
\caption{Influence of the panel position on lifetime, maximum drag and differential drag referred to the panel position at $x=0$, $\alpha_T=0.91$.}
\label{fig:panel_position}
\end{figure}

Taking up the theoretical approach for further increasing the maximum drag, changing the panel position as shown in Fig.~\ref{fig:panpo_cad} poses another possibility to increase the surface perpendicular to the flow, which is marked in orange. This parameter study of the panel position was performed on a main body of geometry option~A, since the geometry option~C has limited width to move the panel due to the sharp nose, and since changing the panel position on geometry option~B does not change the amount of surface elements perpendicular to the flow. The influence of the panel position on lifetime, maximum drag and differential drag can be seen in Fig.~\ref{fig:panel_position} and was estimated using ADBSat, by gradually moving the panel from the centered position to the edge of the frontal circular area. It can be seen, that as expected, the maximum drag can be increased, but simultaneously the lifetime decreases due to the greater $A_{ref}$ in nominal flight configuration. Hence, changing the panel position is not a recommended design variation for the use of differential drag control methods in the given case.
\subsection{Differential lift optimization}
\label{subsec:diff_goal2}
The lift values compared and shown in this section are referring to the optimum angle of attack for the respective satellite geometry. Hence, the compared lift values represent the maximum possible lift, but may be experienced at different AOSs.

\begin{figure}
  \centering
    \includegraphics{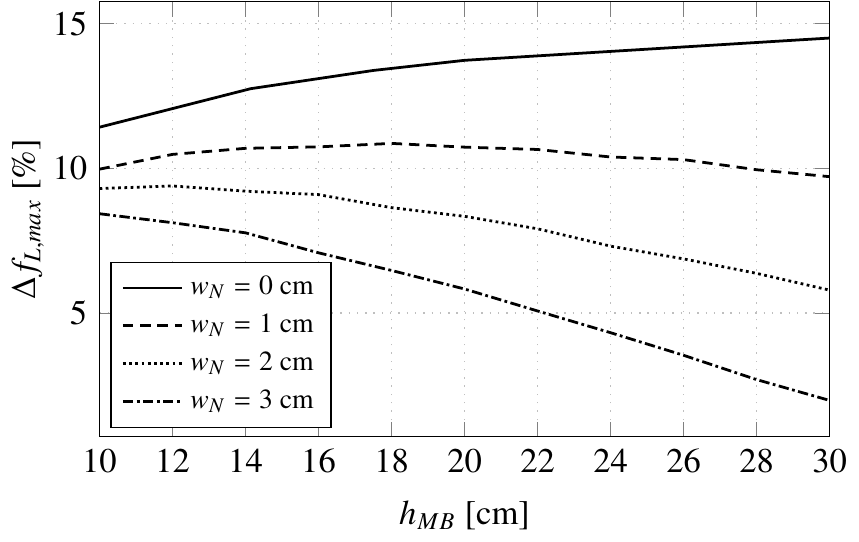}
\caption{Influence of nose width $w_N$ and main body height $h_{MB}$ on the lift force, $\alpha_T=1.00$}
\label{fig:diffuse_lift_nose}
\end{figure}

Optimizing the profile of the satellite using the EPOT's lift optimization function led to a simple geometry as presented in Fig.~\ref{subfig:Design3a} for a given main body height. However, here the side surface of the main body and of the panel form a different angle with the macroscopic velocity vector. Hence, the geometry can be further optimized by increasing the main body height to the original height of the panel as shown in Fig.~\ref{subfig:Design3b}. In order to validate the EPOT's output, other geometries (e.g. the designs built for the investigations within the previous section) have been evaluated as well, but none of them were able to increase the lift such as design Lift~2, when compared to the reference satellite. However, since design Lift~2 contains a sharp nose geometry, possibly posing a problem towards the accommodation of the payload, the consequences of a mandatory frontal area were investigated as well as shown in Fig.~\ref{fig:diffuse_lift_nose}. In order to obtain the ADBSat results for the change in lift force, the utilized satellite models were obtained extruding the with the EPOT optimized profile considering the respective main body heights $h_{MB}$ and nose widths $w_N$. It is apparent, that for a given nose width, the overall frontal area increases with increasing main body height, and thus the lift opposite to the desired lift direction increases as well. Hence, the greater the given nose width, the smaller the optimum main body height and achievable increase in lift force. In order to achieve a high increase in lift force, a frontal nose area should be avoided. If not possible, the main body height has to be chosen according to the applied nose width for a maximum increase in lift force.
\subsection{Discussion on material as design factor}
\begin{figure}
  \centering
    \includegraphics{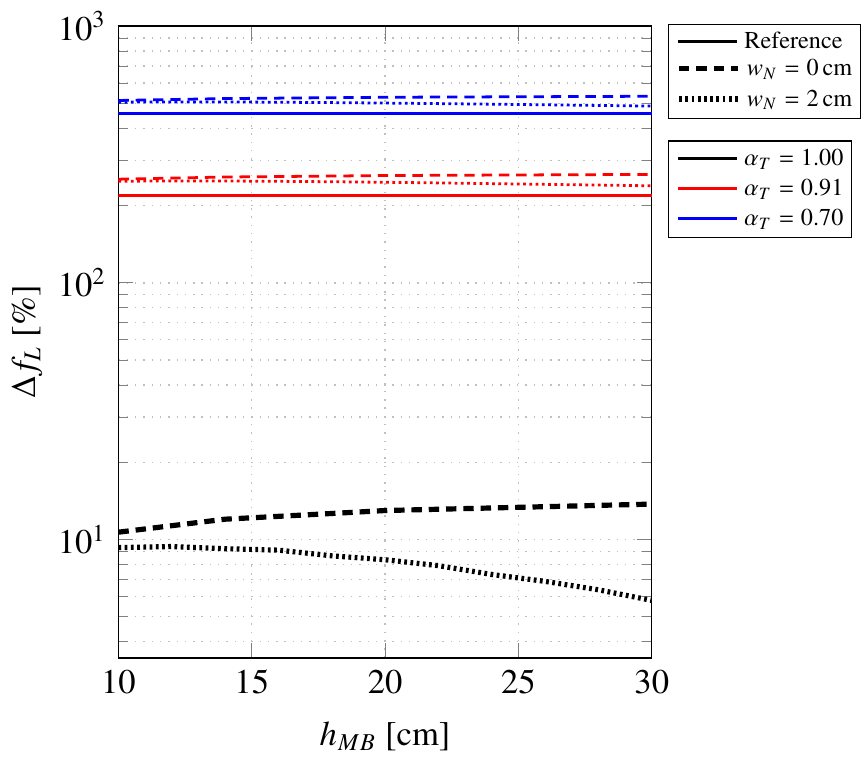}
  \caption{Dependence of lift increase on the main body height and surface properties compared to the reference satellite with $\alpha_T=1.00$.}
  \label{fig:diff_lift_100}
\end{figure}

Within this section, the influence of improved surface properties under the assumption of diffuse re-emission is discussed. In Fig.~\ref{fig:diff_lift_100}, the data obtained for the considerations of a frontal nose area, geometry option~B, and a varying main body height as presented in Section~\ref{subsec:diff_goal2} for different $\alpha_T$ is shown with respect to the data from the reference satellite with $\alpha_T=1.00$. Here, the comparison to the reference satellite of $\alpha_T=1.00$ was chosen in order to visualize the effect of varying $\alpha_T$ and varying the geometry. It can be seen, that changing the reference satellite's $\alpha_T$ from 1.00 to 0.91 increases the lift more than optimizing the geometry with $h_N=0$ for $\alpha_T=1.00$. So not only does the change from diffuse re-emission to specular reflection enable greater atmospheric forces than a design variation, but also a reduction of $\alpha_T$ for diffusely re-emitting materials. Especially for high $\alpha_T$, adapting the material is therefore a more powerful design variation than optimizing the geometry with regard to increased control forces. As Fig.~\ref{fig:diff_lift_100} shows, the effect of a change in geometry (e.g. from $h_N=0$~cm to $h_N=2$~cm) increases with decreasing $\alpha_T$.

\subsection{Performance of the optimized designs}
\label{sec:Simulation_Results}
The final design recommendation for the assumption of a diffuse re-emission is presented in the following and its actual performance was evaluated using PICLas.
\subsubsection{Differential drag and lift - separate consideration}
\label{subsec:diff_lift}
After investigating several design derivations of the reference satellite, a satellite geometry with a revolved lifetime-optimized 2D profile including an additionally added rasterized cross-section swept along the 2D profile as a main body, and an extruded lifetime-optimized 2D
profile for the panel was shown to be advantageous for decreased minimum drag and increased differential drag. The design (Raster 1) is shown in Fig.~\ref{subfig:Design1a} with 2D profiles optimized for $\alpha_T=1.00$.

To increase the experienced lift force in the desired direction, it was shown to be advantageous to increase the side surface area experiencing lift, as well as decreasing the frontal area, which otherwise experiences a lift force in the opposite direction than desired. The design recommendation (Lift~2), which applies for all the tested $\alpha_T$ within the diffuse re-emission section, is shown in Fig.~\ref{subfig:Design3b}.  
\subsubsection{Differential drag and lift}
The recommendation for differential drag and the recommendation for differential lift do not share many geometric characteristics. Since a sustainable operation in VLEO is the goal, design Lift~2 is not suitable for the general use for the relative motion control by means of the atmospheric forces. Only for $\alpha_T=0.70$, no loss in lifetime compared to the reference satellite with $\alpha_T=0.70$ is achieved with design Lift~2. However, another solution for materials with higher $\alpha_T$ is of interest. 

\begin{figure*}
\centering
  \subfloat[Reference satellite - nominal flight configuration\label{subfig:flowfield_diff_00_def}]{%
    \resizebox{0.456\textwidth}{!}{\includegraphics{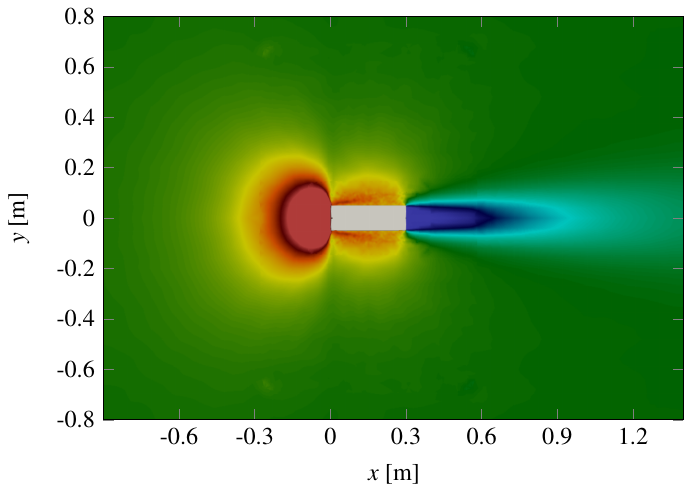}}%
  } \quad
  \subfloat[Recommended design Raster 1 - nominal flight configuration\label{subfig:flowfield_diff_00_rec}]{%
    \resizebox{0.51\textwidth}{!}{\includegraphics{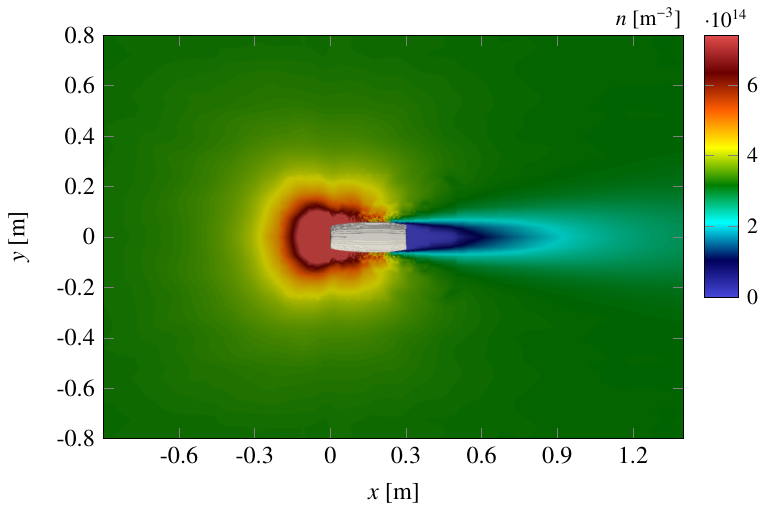}}%
      }
      
  \subfloat[Reference satellite - maximum drag configuration\label{subfig:flowfield_diff_90_def}]{%
    \resizebox{0.456\textwidth}{!}{\includegraphics{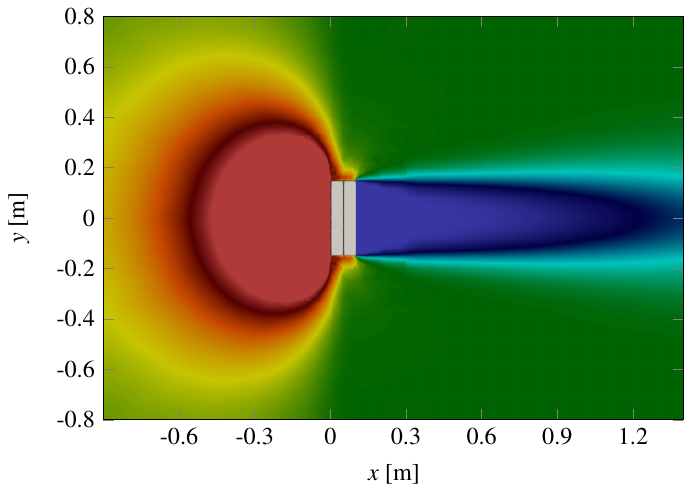}}%
  } \quad
  \subfloat[Recommended design Raster 1 - maximum drag configuration\label{subfig:flowfield_diff_90_rec}]{%
    \resizebox{0.51\textwidth}{!}{\includegraphics{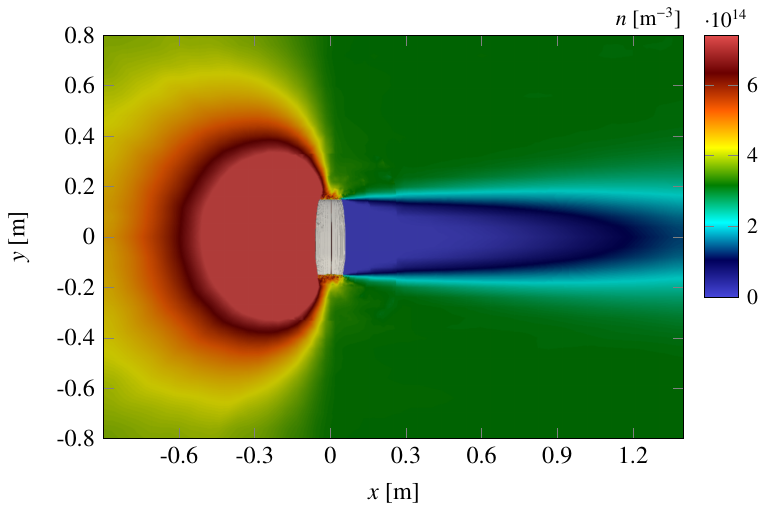}}%
  }
  
  \subfloat[Reference satellite - maximum lift configuration\label{subfig:flowfield_diff_45_def}]{%
    \resizebox{0.456\textwidth}{!}{\includegraphics{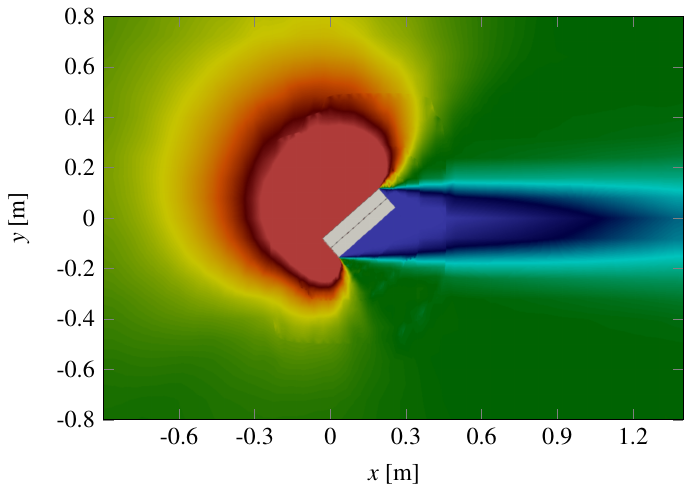}}%
  } \quad
  \subfloat[Recommended design Raster 1 - maximum lift configuration\label{subfig:flowfield_diff_45_rec}]{%
    \resizebox{0.51\textwidth}{!}{\includegraphics{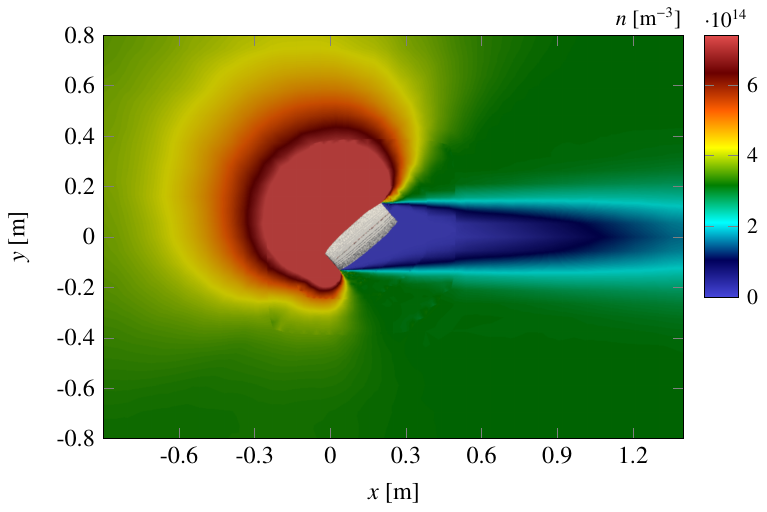}}%
  }
\caption{Simulated particle density $n$ in the flow-field around the reference satellite and the recommended design Raster 1 for $\alpha_T=1.00$.}
\label{fig:flowfield_diff}
\end{figure*}

Raster~1 was shown to be the best trade-off, since the vertical side surfaces for increasing the maximum drag simultaneously have a positive effect on increasing the maximum lift force and therefore, an improvement in all critical parameters was achieved. The flow-fields in all three flight configurations for the reference satellite and design Raster~1 are given in Figs.~\ref{subfig:flowfield_diff_00_def}-\ref{subfig:flowfield_diff_45_rec}. It can be seen, that the 2D profile optimization of main body and panel reduces the shock occurring in Fig.~\ref{subfig:flowfield_diff_00_rec} for $x<0$~m. The reduction of opposed lift experienced by the frontal area is visible in Fig.~\ref{subfig:flowfield_diff_45_rec}, where the shock preceding the frontal area of the recommended design is smaller compared to the reference satellite. The respective specific forces and lifetimes for the designs Raster~1 and Lift~2 are given in Tab.~\ref{tab:summary_rec_designs}. Due to the symmetry, the minimum lift is equal to zero. Since the altitude loss per maneuver plays an important role, the value for drag in the maximum lift configuration $f_{D,L}$ is given as well.

\begin{table}
	\centering
		\begin{tabular}{ll|c|c|c} 
		& & Reference & Raster 1 & Lift 2 \\ \hline
		$\Delta t_l$ &\%& - &   +12.26  & * -8.14 \\
		$\Delta f_{\Delta D}$ &\% & - &  +2.44  &  * -4.17 \\
		$\Delta f_{\Delta L}$ &\% & - & +3.14  & +16.03  \\ \hline
        $t_L$ &d & 157.09 & 176.35 & * 142.27\\
		$|f_{\Delta D}|$ &\si{\metre\per\square\second}& 8.303$\cdot 10^{-6}$  & 8.508$\cdot 10^{-6}$  & * 8.220$\cdot10^{-6}$ \\
		$|f_{\Delta L}|$ &\si{\metre\per\square\second}& 6.221$\cdot10^{-7}$ & 6.416$\cdot10^{-7}$ & 7.218$\cdot10^{-7}$ \\ \hline
		$|f_{D,min}|$ &\si{\metre\per\square\second}& 2.157$\cdot10^{-6}$  & 1.922$\cdot10^{-6}$  &* 2.380$\cdot10^{-6}$  \\
		$|f_{D,max}|$ &\si{\metre\per\square\second}& 1.046$\cdot10^{-5}$  & 1.043$\cdot10^{-5}$  &* 1.060$\cdot10^{-5}$  \\
		$|f_{L,max}|$ &\si{\metre\per\square\second}& 3.110$\cdot10^{-7}$ & 3.208$\cdot10^{-7}$  & 3.609$\cdot10^{-7}$   \\
		$|f_{D,L}|$ &\si{\metre\per\square\second}& 7.721$\cdot10^{-6}$ & 7.362$\cdot10^{-6}$  & 7.025$\cdot10^{-6}$ \\ 
	\end{tabular}
	\caption{Data for the reference satellite, design Raster 1 and Lift 2, $\alpha_T=1.00$. Data obtained using PICLas, with * marked data obtained using ADBSat.}
	\label{tab:summary_rec_designs}%
\end{table}

\section{Design optimization - specular reflection}
\label{sec:process2}
The satellite models within this section were generally evaluated with the Cercignani-Lampis-Lord (CLL) model of ADBSat. As the required input parameters are $\alpha_n$ and $\sigma_t$ and not $\alpha_T$, they were derived from the chosen $\alpha_T$ given in Tab.~\ref{tab:surface_properties}, as presented in \ref{sec:DerivationADBSat}.
\subsection{Differential drag optimization}
Within this section, the influence of general shape, tail geometry, main body height, and frontal area on the experienced drag forces is presented. Sections~\ref{subsubsec:mainbodyheight} and \ref{subsubsec:frontalarea} differ from the respective sections under the design optimization process of diffuse re-emission, as they are linked to the results of the investigation of general shape and tail geometry.
\subsubsection{General shape and tail geometry}
\label{subsec:spec_shape}
\begin{figure*}
\centering
\begin{subfigure}{.36\textwidth}
  \centering
    \includegraphics{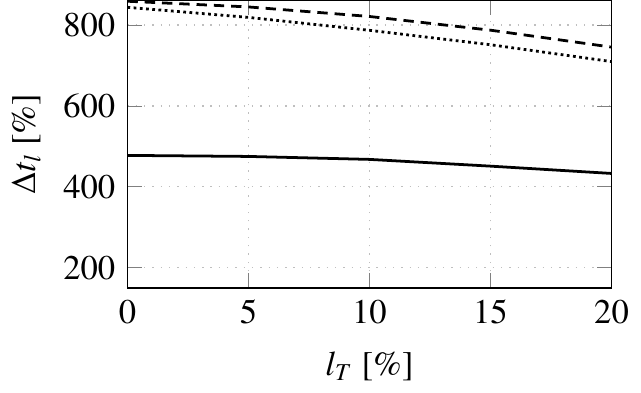}
  \caption{$\alpha_n=0.00\quad\sigma_t=0.00$}
  \label{subfig:spec_tail_000}
\end{subfigure}%
\begin{subfigure}{.315\textwidth}
  \centering
    \includegraphics{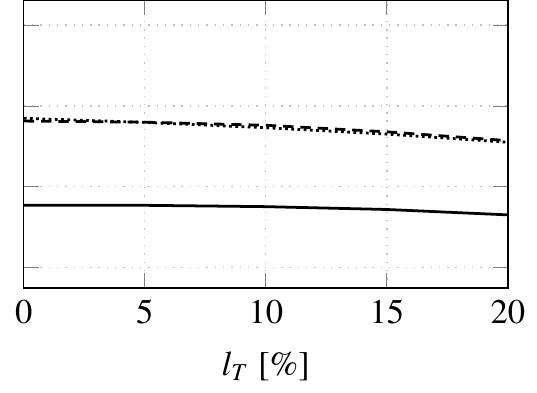}
  \caption{$\alpha_n=0.09\quad\sigma_t=0.0459$}
  \label{subfig:spec_tail_009}
\end{subfigure}
\begin{subfigure}{.315\textwidth}
  \centering
    \includegraphics{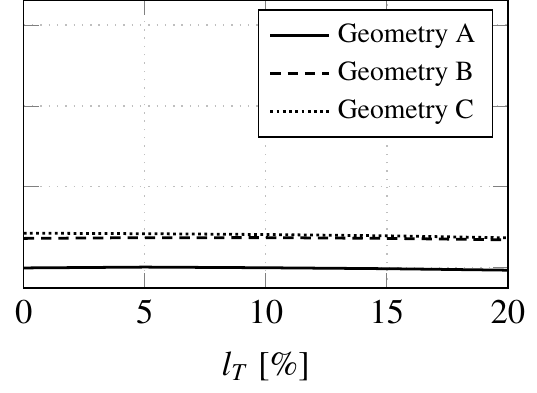}
  \caption{$\alpha_n=0.30\quad\sigma_t=0.1627$}
  \label{subfig:spec_tail_030}
\end{subfigure}
\caption{Increase in lifetime $\Delta t_L$ depending on general shape and tail length $l_T$.}
\label{fig:spec_tail}
\end{figure*}

In a first step, the best possible volume derivation of the optimized 2D profile and the appropriate tail length (if any) were investigated. An overview of the lifetime in the nominal flight configuration influenced by the three geometry derivation options A, B and C and the tail length is given in Fig.~\ref{fig:spec_tail}. The data shown was obtained using ADBSat. 
For very low energy accommodation coefficients, the geometry option B is the most advantageous for a high increase in lifetime. With increasing energy accommodation coefficient, geometry option C is the more promising volume derivation option. From Fig.~\ref{subfig:spec_tail_000} and Fig.~\ref{subfig:spec_tail_009} it can be seen, that for the mentioned advantageous volume derivation options the design with no tail geometry is the most promising design respectively. 
\subsubsection{Main body height}
\label{subsubsec:mainbodyheight}
\begin{figure*}
\centering
\begin{subfigure}[b]{.33\textwidth}
  \centering
    \includegraphics{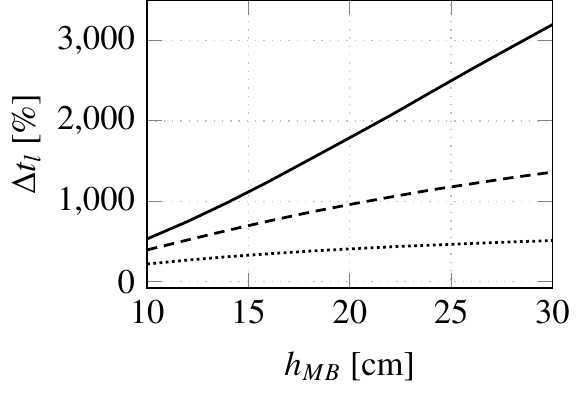}
  \caption{Lifetime}
  \label{subfig:spec_height_lifetime}
\end{subfigure}%
\begin{subfigure}[b]{.33\textwidth}
  \centering
    \includegraphics{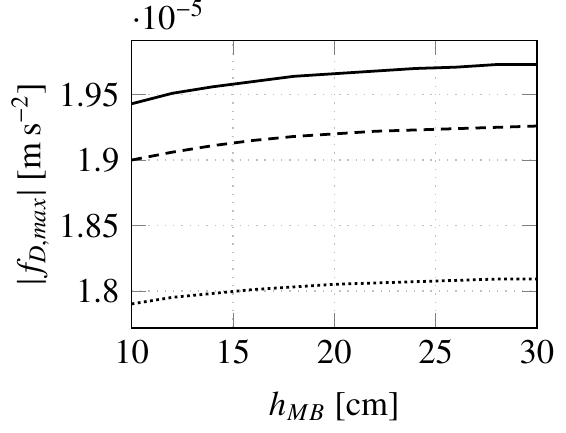}
  \caption{Maximum Drag}
  \label{subfig:spec_height_maxdrag}
\end{subfigure}
\begin{subfigure}[b]{.31\textwidth}
  \centering
    \includegraphics{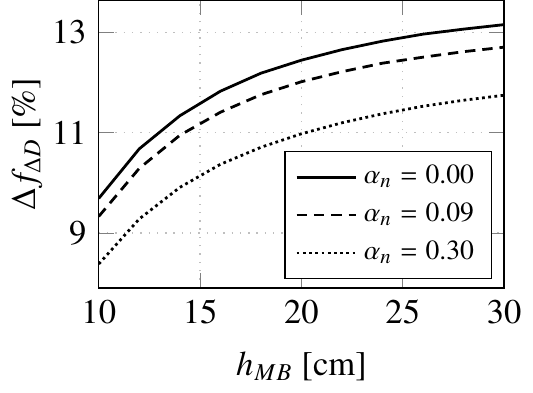}
  \caption{Differential Drag}
  \label{subfig:spec_height_diffdrag}
\end{subfigure}
\caption{Increase in lifetime, drag, and differential drag depending on the main body height $h_{MB}$.}
\label{fig:spec_height}
\end{figure*}

The volume derivation option~B is promising regarding a long lifetime of the satellite. Since the original 2D optimization tool does not investigate different main body heights for option~B, their influence was investigated further. Figure~\ref{fig:spec_height} gives an overview of the lifetime in nominal flight configuration, maximum drag in maximum drag configuration and differential drag depending on the main body height. The profiles used were obtained utilizing the EPOT with a given extrusion height, which is set equal to the main body height $h_{MB}$. The data shown was obtained using ADBSat.
It can be seen, that especially for very low energy accommodation coefficients, a high main body height has great potential for increasing the overall lifetime. Additionally, an increased main body height is advantageous for increasing the experienced drag in maximum drag configuration and thus also advantageous for a high differential drag force. The with regard to a long lifetime and high differential drag optimized satellite, obtained using geometry option B, is depicted as design SpecOpt in Fig.~\ref{subfig:Design4a}.
\subsubsection{Consequences of practical considerations}
\label{subsubsec:frontalarea}
As aerodynamically promising design~SpecOpt is, the practicability however of accommodating the payload can be lowered due to the sharp nose. A more practicable solution would be to include a frontal area. Therefore, the following nose geometries were considered: a constant nose width of $w_N=3$~cm, a constant nose area equal to a third of the reference satellite's frontal area $A_{N,0}$, a constant nose area equal to the half of $A_{N,0}$ as well as a constant nose area equal to two thirds of $A_{N,0}$. An exemplary depiction of the tested satellite geometries can be found as design SpecPrac in Fig.~\ref{subfig:Design4b}. 

\begin{figure}
\centering
\begin{subfigure}[t]{.48\textwidth}
  \centering
    \includegraphics{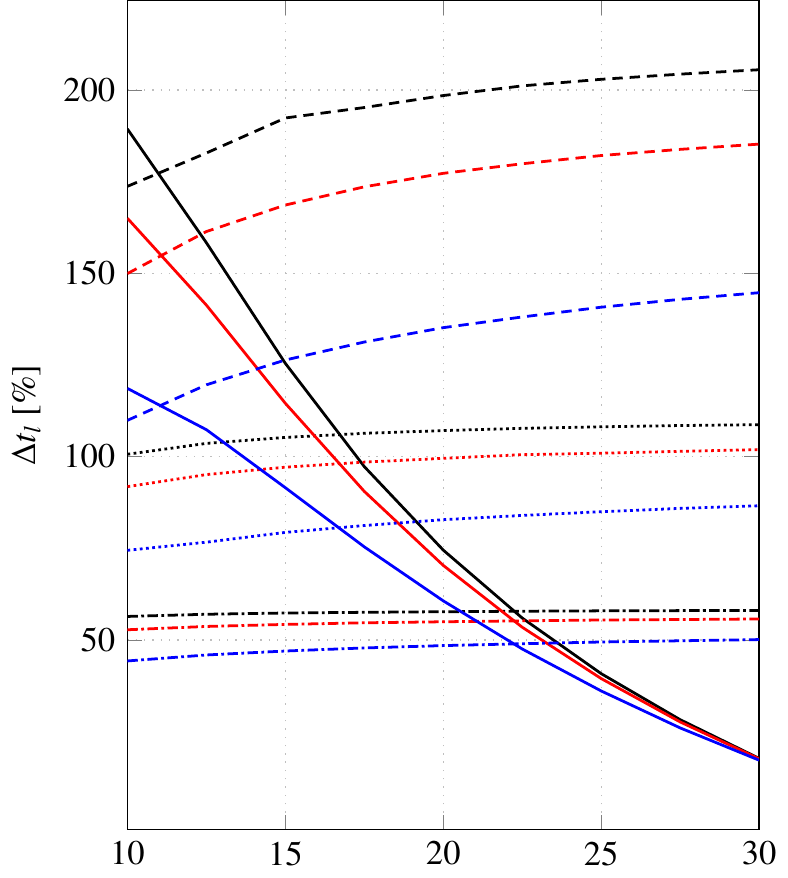}
  \caption{Lifetime}
  \label{subfig:spec_nose_lifetime}
\end{subfigure}

\begin{subfigure}[t]{.48\textwidth}
  \centering
    \includegraphics{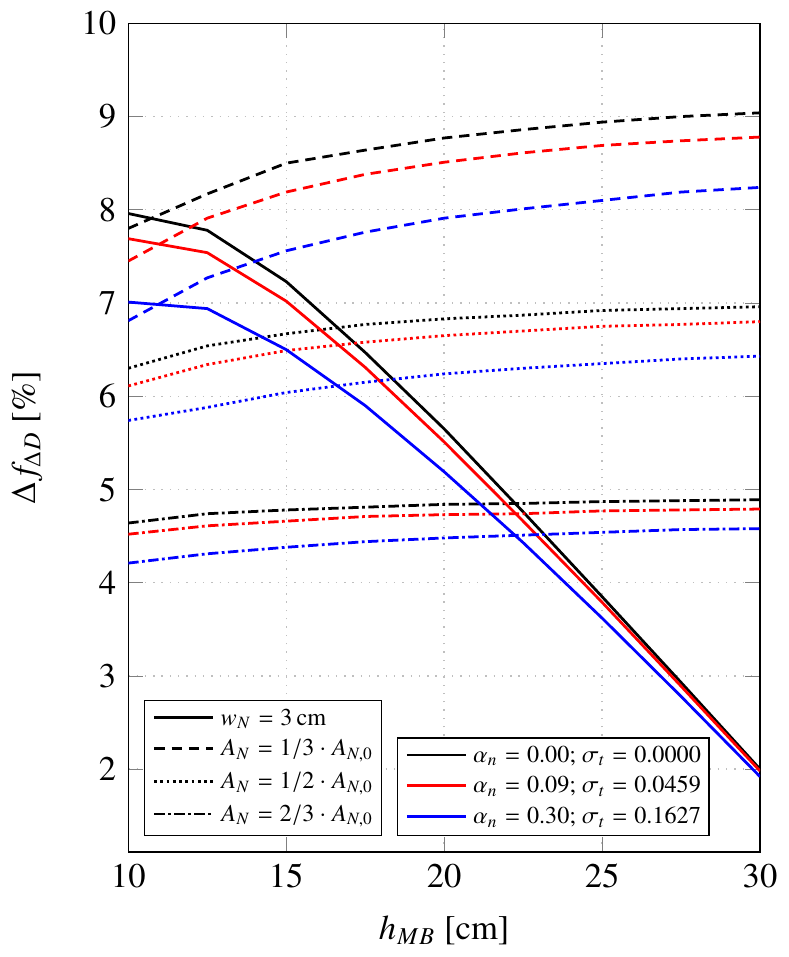}
  \caption{Differential Drag}
  \label{subfig:spec_nose_drag}
\end{subfigure}
\caption{Lifetime and differential drag depending on the main body height for different frontal areas.}
\label{fig:spec_nose_drag_lifetime}
\end{figure}

The dependence of lifetime in nominal flight configuration and differential drag for the considered nose geometries and different $\alpha_T$ is shown in Fig.~\ref{fig:spec_nose_drag_lifetime}. The utilized satellite models were derived using the EPOT considering $h_{MB}$ and $w_N$, and the data shown was obtained using ADBSat. It can be seen that for a given $w_N$ regardless of $h_{MB}$, a smaller main body height is more advantageous regarding the lifetime, as an increase in $h_{MB}$ simultaneously increases the frontal area and therefore the drag. If the frontal area remains constant for different $h_{MB}$, i.e. a decreasing nose width $w_N$ for increasing main body height $h_{MB}$, the influence of $\alpha_n$ and $\sigma_t$ as well as the achievable improvement in lifetime and differential drag decreases with increasing frontal area. Therefore it is recommended to choose a main body height as small as possible for a mandatory given nose width. However, if the value of the nose width is not of direct importance, but a frontal area is required, it is recommended to keep the nose width as small as possible. As visible when comparing the scales of the y-axes of Fig.~\ref{subfig:spec_height_lifetime} and Fig.~\ref{subfig:spec_nose_lifetime}, any type of frontal area leads to a significant loss in lifetime and thereby also reduces the achievable increase in differential drag compared to the reference satellite.
\subsection{Differential lift optimization}
\begin{figure}
  \centering
    \includegraphics{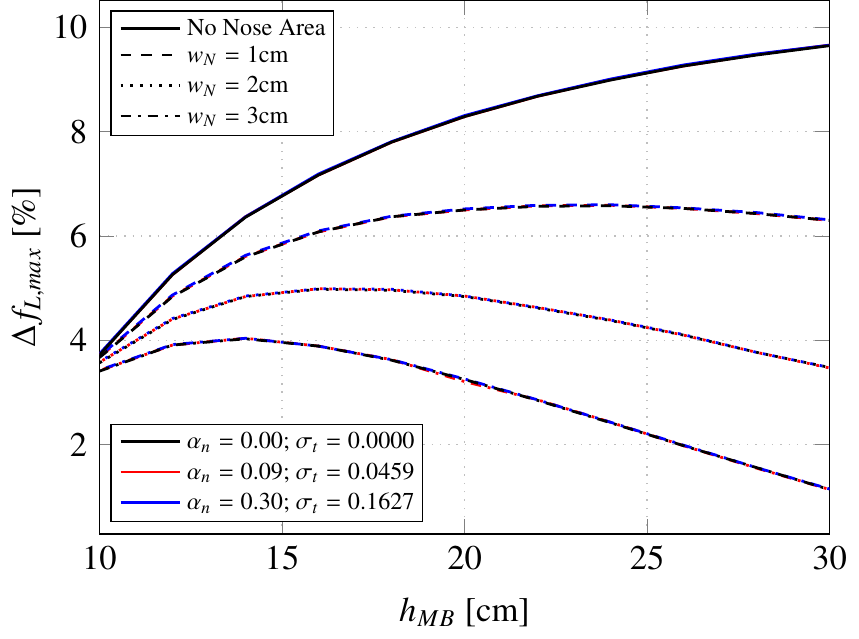}
\caption{Influence of the main body height $h_{MB}$ and different nose widths on the lift.}
\label{fig:spec_nose_lift}
\end{figure}

The suggested profile by the 2D optimization tool is equal to the one as derived in Section~\ref{subsec:diff_goal2} and thus, the influence of different frontal areas was investigated again for the respective $\alpha_T$. Therefore, the satellite models from Section~\ref{subsec:diff_goal2} were evaluated for $\alpha_T=[0.00;0.09;0.30]$ using ADBSat. Figure~\ref{fig:spec_nose_lift} shows, that whereas the absolute value of specific lift force also depends on the energy accommodation coefficient, no dependence of the percentage increase in lift with respect to the reference satellite can be seen. Additionally it is visible that the optimal main body height is depending on the nose width similarly to the equivalent investigation results in the diffuse re-emission section. For a high increase in experienced lift force due to solely geometry design factors, a frontal area should be avoided or, if not possible, the nose width should be kept as small as possible.

\subsection{Performance of the optimized designs}
\label{sec:results2}
The final design recommendation for the assumption of a specular reflection is presented in the following and its actual performance was evaluated using PICLas.
\subsubsection{Differential drag and lift - separate consideration}
Especially the extruded optimized 2D profile with an extrusion height equal to the main body height of 30~cm was shown to drastically increase the lifetime and thus the differential drag for low $\alpha_T$. The resulting design (SpecOpt) is shown in Fig.~\ref{subfig:Design4a} with a 2D profile optimized for $\alpha_T=0.00$. This design is presented to increase the understanding of advantageous design characteristics prior to deriving a realistic design recommendation.

Similar to the case of diffuse re-emission, it was shown to be advantageous to increase the total side surface area by inclination, as well as eliminating a frontal area. Regardless of the tested GSIM and $\alpha_T$, the design optimal for differential lift is given in Fig.~\ref{subfig:Design3b}, as presented in Section~\ref{subsec:diff_goal2}.
\subsubsection{Differential drag and lift}
Opposite to the designs within the section for diffuse re-emission, the two previous satellite design recommendations share many geometry characteristics. The only difference between these two is the slightly curved profile, which is the result of the 2D optimization tool with regard to an increased lifetime.  As design SpecOpt has a higher increase in lifetime (first priority) compared to design Lift~2, but almost the same increase in lift (second priority), design SpecOpt is also well suited for the use of both, differential lift and drag control methods. However, while design SpecOpt is the theoretical optimal design for the use in VLEO in a satellite formation, it not only poses difficulties for the accommodation of the payload, but also for other factors such as manufacturing or attitude control. Hence, a more practicable but less optimal solution is presented and discussed. A minimum required frontal area is assumed with a nose width of $w_N=3$~cm\footnote{This value was adopted to primarily study performance impacts. It remains to be decided which value is reasonable.}. In order to keep the loss in lifetime as small as possible, the main body height should remain as small as possible as well and therefore it is set equal to the reference satellite with $h_{MB}=10$~cm. The new design (SpecPrac) is depicted in Fig.~\ref{subfig:Design4b} with a 2D profile optimized for $\alpha_T=0.00$. It was obtained using the EPOT geometry B under consideration of given $w_N$ and $h_{MB}$. The panel is optimized for an increased lifetime as well using geometry option B and a given total height of $h_{max}=30$~cm.

\begin{figure*}
\centering
  \subfloat[Reference satellite - nominal flight configuration\label{subfig:flowfield_spec_00_def}]{%
    \resizebox{0.456\textwidth}{!}{\includegraphics{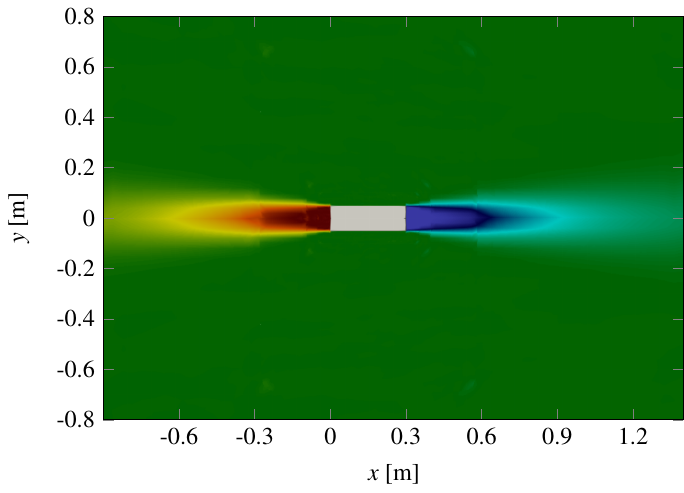}}%
  } \quad
  \subfloat[Practical design SpecPrac - nominal flight configuration\label{subfig:flowfield_spec_00_prac}]{%
    \resizebox{0.51\textwidth}{!}{\includegraphics{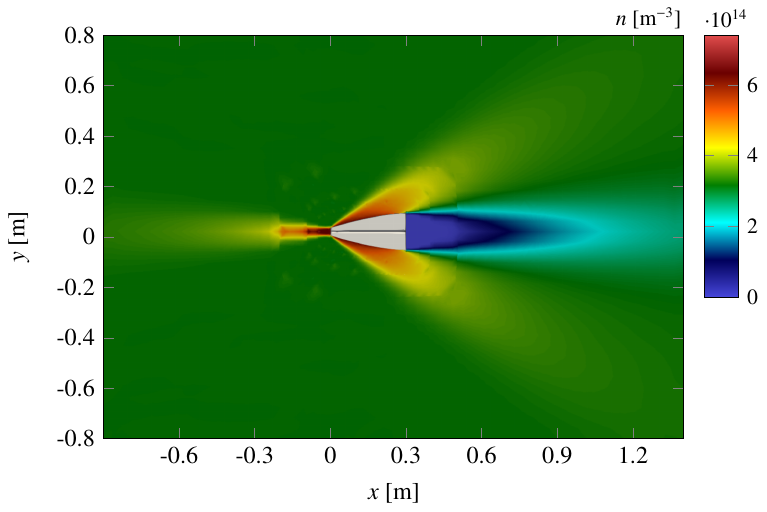}}%
      }
      
  \subfloat[Reference satellite - maximum drag configuration\label{subfig:flowfield_spec_90_def}]{%
    \resizebox{0.456\textwidth}{!}{\includegraphics{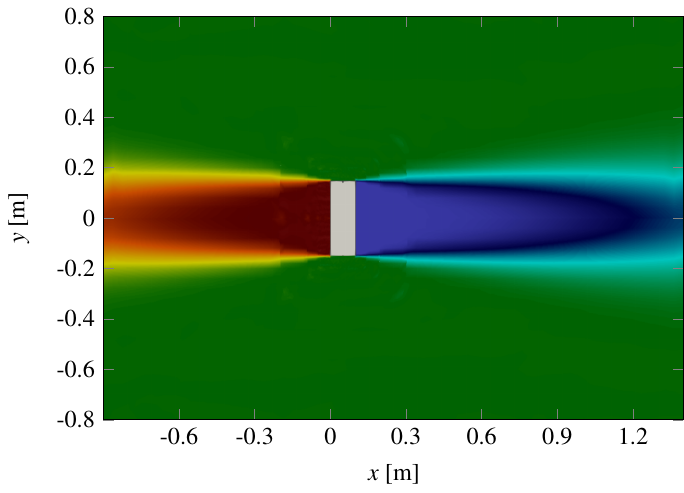}}%
  } \quad
  \subfloat[Practical design SpecPrac - maximum drag configuration\label{subfig:flowfield_spec_90_prac}]{%
    \resizebox{0.51\textwidth}{!}{\includegraphics{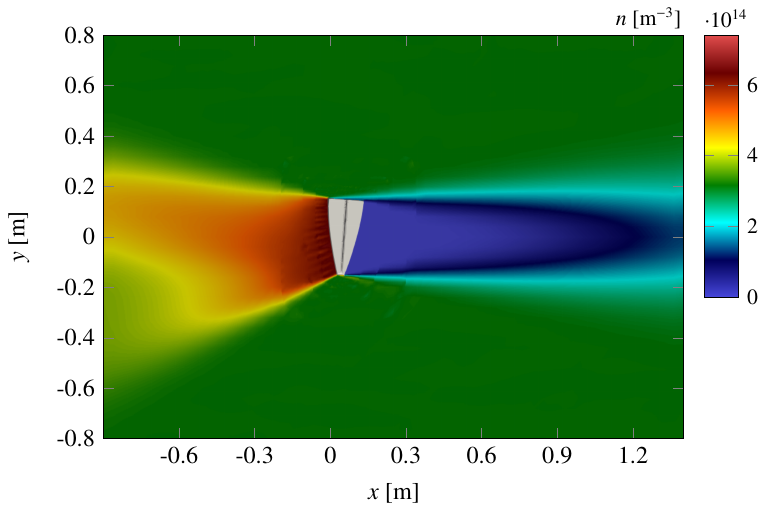}}%
  }
  
   \subfloat[Reference satellite - maximum lift configuration\label{subfig:flowfield_spec_45_def}]{%
    \resizebox{0.456\textwidth}{!}{\includegraphics{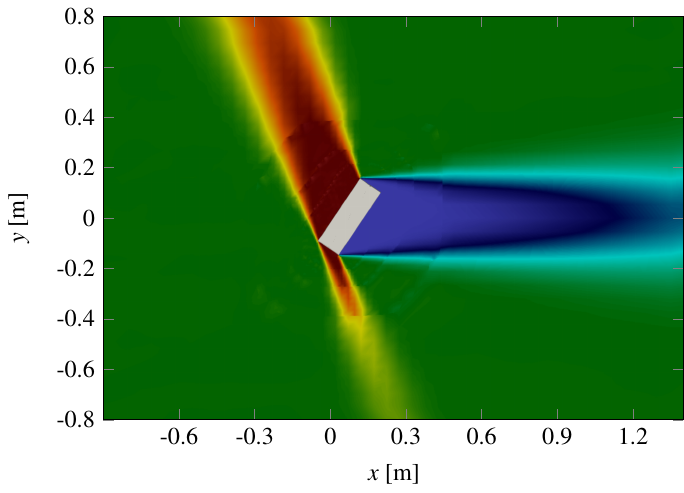}}%
  } \quad
  \subfloat[Practical design SpecPrac - maximum lift configuration\label{subfig:flowfield_spec_45_prac}]{%
    \resizebox{0.51\textwidth}{!}{\includegraphics{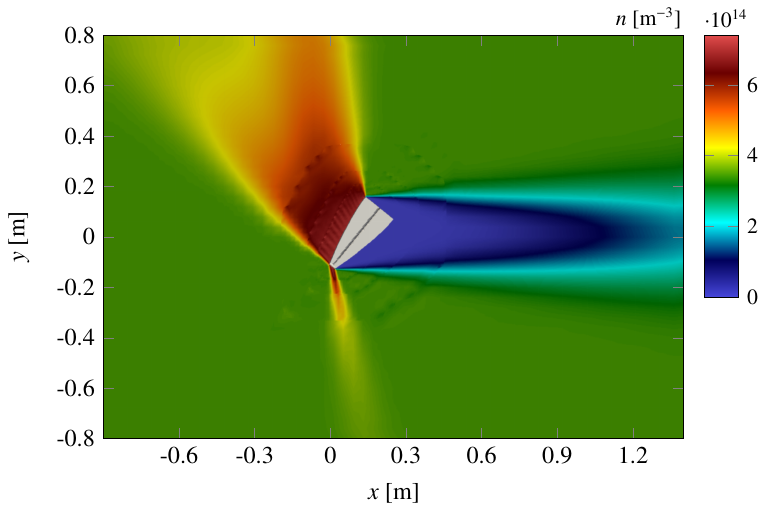}}%
  }
\caption{Simulated particle density $n$ in the flow-field around the optimal design SpecOpt and the more practicable version SpecPrac for $\alpha_T=0.00$.}
\label{fig:flowfield_spec_prac}
\end{figure*}

The according flow-fields for the reference and the design SpecPrac are given in Fig.~\ref{fig:flowfield_spec_prac}. In Tab.~\ref{tab:summary_rec_designs_spec}, the corresponding data is given. The lifetime decreasing effect of a frontal area in the case of the practicable design recommendation can be seen in Fig.~\ref{subfig:flowfield_spec_00_prac}, where the deflection of the impinging particles by 180° leads to an accumulation of particles ahead of the satellite. However, the reduction of the total frontal area compared to the reference led to an increase in lifetime. In Fig.~\ref{subfig:flowfield_spec_90_prac} the particles are reflected in a wider range with an (in the picture) downward shift due to the increased curvature of the main body's side surface. The overlapping region of particles reflected by the panels and particles reflected by the rearward part of the side surface can be seen in the range around $-0.5$~m~$<\, y\, <0.3$~m. In this case, the non-symmetrical deflection in the maximum drag configuration leads to a lift force, which is comparatively small, but nonetheless can build up to a considerable effect over time. As the main body's frontal surface area is smaller than the reference's, it was possible to increase the lift force (see Fig.~\ref{subfig:flowfield_spec_45_prac}).
\begin{table}
	\centering
		\begin{tabular}{ll|c|c|c} 
		  && Reference & 4b & 3b \\ \hline
		$\Delta t_l$& \%& - &   +200  & * +2,464 \\
		$\Delta f_{\Delta D}$ &\% & - &  +4.01  &   * +13.52\\
		$\Delta f_{\Delta L}$ &\% & - & +0.66  & +10.03  \\ \hline
        $t_L$& d & 142.40 & 427.14 & * 3745.19\\
		$|f_{\Delta D}|$ &\si{\metre\per\square\second}& 1.722$\cdot 10^{-5}$  & 1.792$\cdot 10^{-5}$  & * 1.973$\cdot10^{-5}$ \\
		$|f_{\Delta L}|$ &\si{\metre\per\square\second}& 1.380$\cdot10^{-5}$ & 1.389$\cdot10^{-5}$ & 1.518$\cdot10^{-5}$ \\ \hline
		$|f_{D,min}|$ &\si{\metre\per\square\second}& 2.380$\cdot10^{-6}$  & 7.934$\cdot10^{-7}$  &* 0.905$\cdot10^{-7}$  \\
		$|f_{D,max}|$ &\si{\metre\per\square\second}& 1.960$\cdot10^{-5}$  & 1.871$\cdot10^{-5}$  &* 1.982$\cdot10^{-5}$  \\
		$|f_{L,max}|$ &\si{\metre\per\square\second}& 6.900$\cdot10^{-6}$ & 6.945$\cdot10^{-6}$  & 7.592$\cdot10^{-6}$   \\
		$|f_{D,L}|$ &\si{\metre\per\square\second}& 1.157$\cdot10^{-5}$ & 1.064$\cdot10^{-5}$  & 7.224$\cdot10^{-6}$ \\ 
	\end{tabular}
	\caption{Data for the reference satellite, design SpecOpt and Lift 2, $\alpha_T=0.00$. Data obtained using PICLas, with * marked data obtained using ADBSat.}
	\label{tab:summary_rec_designs_spec}%
\end{table} 

\section{Conclusion}
In this work, the design of a given reference satellite has been optimized with regard to the use of differential lift and drag control methods and a sustainable application in orbit. The influence of design variations was generally evaluated using ADBSat, and the performance of the promising designs was verified using the DSMC code PICLas. 

For materials with diffuse re-emission, the design optimized for differential drag and the design optimized for differential lift did not share many geometric characteristics. Hence, when deriving a design optimized for both differential lift and drag, a priority of either lift or drag has to be set. However, for all given surface properties, an optimization with regard to the use of differential lift and drag control methods was possible with the primary objective of increasing the lifetime. Even for the case of complete accommodation $\alpha_T=1.00$, differential drag and lift could be increased by 2~\% and 3~\% respectively, while at the same time increasing the lifetime under the set conditions by 12~\%. Despite the possible increase of differential forces, the differential lift remains two orders of magnitude smaller compared to the achievable differential drag. Hence, a differential lift maneuver necessary for influencing the out-of-plane motion cannot be performed without a significant loss in altitude. Generally, the design variations had more effect for specular materials than for materials reflecting particles diffusely. However, the theoretical optimal design for a satellite with specular materials entails little practicability. A more practicable solution was presented, which still achieved an increased lifetime of 200~\%, increased differential drag by 4~\% and differential lift by 1~\% compared to the reference satellite with the same surface properties. The same optimized design compared to the reference satellite with a diffuse re-emitting material achieved an increase of 172~\% in lifetime, 116~\% in differential drag and 2133~\% in differential lift.  Therefore, the more promising approach for realizing relative motion control in a satellite formation by differential lift and drag for high $\alpha_T$ is an improvement of the currently available surface properties, as only for specularly reflecting materials the differential lift and drag are of same order of magnitude. 

However, the optimized 3D designs presented and discussed throughout the article are derived from optimal 2D profiles and thus, the direct optimization of the 3D bodies for possibly further increasing the performance is currently still pending. Additionally, the combination of different surface properties in one satellite design was not considered and represents a possible continuation of this work. Applying a diffusely reflecting frontal area and specular side surfaces for a combination of sustainable application in orbit and increased forces for the use in relative motion control within the formation is one example of the possibilities for mixed surface properties. The improvement of the materials is the most influential design parameter in order to make the sustainable implementation of thrust-free control methods possible and represents an important step towards future space systems. Hence, advancing material research is of great interest for the application of VLEO satellites and satellite formations. 
\appendix
\section{Comparison of ADBSat and PICLas results}
\label{sec:PIClasADBSat}
\begin{figure*}
    \includegraphics{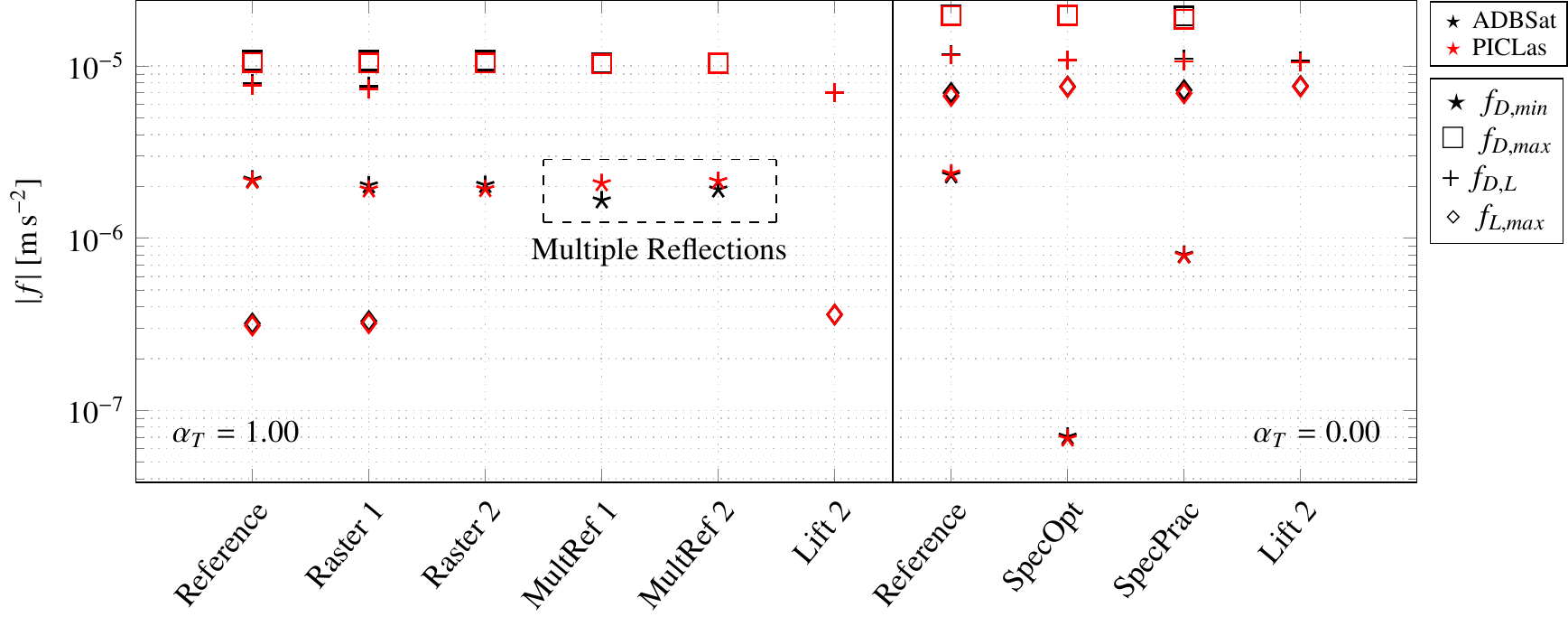}
  \caption{Comparison of ADBSat and PICLas results.}
  \label{fig:ADBSatComp}
\end{figure*}
In this section, the results obtained using ADBSat and PICLas for the recommended designs presented in the previous sections are compared. Figure~\ref{fig:overview} gives an overview over the designs whose results are listed in Fig.~\ref{fig:ADBSatComp}.

It can be seen, that the results of ADBSat generally comply with the ones obtained using PICLas. The biggest difference is visible for design MultRef 1 and 2 for the minimum drag. However, design MultRef 1 and 2 are the designs created in order to promote multiple reflections. Here, the minimum drag obtained using PICLas is 25~\% higher for design MultRef 1 and 12~\% higher for design MultRef 2 than estimated with ADBSat. 

If only geometries suitable for the panel method are regarded, the values for the specific forces obtained using PICLas are generally smaller than the ones calculated with ADBSat. However, the biggest difference is only a 4~\% smaller maximum drag of PICLas compared to ADBSat. The generally smaller specific forces within PICLas can be explained due to small differences in the edges of the models. The surface mesh used within ADBSat is better suited to model complex surfaces as the triangular shape of the mesh elements allows more narrow and sharp edges. On the other side, sharp edges or small radii within the mesh for PICLas can pose a problem for the 3D mesh generation. Here, rounded and sharp edges are approximated with a greater radius, as a smaller radius increases the cells and thereby the necessary amount of simulated particles and thus simulation time. Altogether it can be said, that for geometries without multiple reflections the results of ADBSat and PICLas do not differ more than 4~\% and the differences can be traced back to differences in the mesh.

\section{Derivation of the ADBSat input parameters}
\label{sec:DerivationADBSat}
For the application of the CLL model in ADBSat, the normal energy accommodation coefficient~$\alpha_n$ and the tangential momentum accommodation coefficient~$\sigma_t$ are required and therefore are derived from the given $\alpha_T$ in Tab.~\ref{tab:surface_properties}. The following derivation is based on   transformations of Sentman's equation by Hild \cite{Hild.2021} and the consideration of a 2D profile.

Since the kinetic energy $E=\frac{1}{2}mv^2$ is proportional to the squared momentum $I^2=m^2v^2$ under the assumption of constant mass, Eq. \ref{eq:at} can be also expressed as
\begin{equation}
\label{eq:at_impulse}
\alpha_T = \frac{I_i^2 - I_r^2}{I_i^2-I_w^2},
\end{equation}
with the momentum $I$ consisting of its normal and tangential components $p$ and $\tau$
\begin{equation}
\label{eq:impulse}
I^2=p^2+\tau^2.
\end{equation}
The normal and tangenatial energy accommodation coefficient $\alpha_n$ and $\alpha_t$ hence can be defined as
\begin{equation}
\label{eq:alpha_n}
\alpha_n=\frac{p_i^2-p_r^2}{p_i^2-p_w^2},
\end{equation}
\begin{equation}
\label{eq:alpha_t}
\alpha_t=\frac{\tau_i^2-\tau_r^2}{\tau_i^2-\tau_w^2}. 
\end{equation}
The parameter $\tau_i$ and $p_i$ as properties of the impinging particle depend on the properties of the atmosphere and are proportional to the following terms \cite{Hild.2021}:
\begin{equation}
\label{eq:tau_i}
\tau_i \propto \epsilon \left[ \gamma(1+\text{erf}(\gamma s))+\frac{1}{s\sqrt{\pi}}e^{-\gamma^2s^2} \right],
\end{equation}
\begin{equation}
\label{eq:p_i}
p_i \propto \gamma \left[ \gamma (1+\text{erf}(\gamma s))+\frac{1}{s\sqrt{\pi}}e^{-\gamma^2s^2} \right] + \frac{1}{2s^2}(1+\text{erf}(\gamma s)).
\end{equation}
$\tau_w$ is per definition equal to zero, and the normal momentum $p_w$ carried away after reaching thermal equilibrium with the wall is proportional to
\begin{equation}
\label{eq:p_w}
p_w \propto \frac{1}{2} \sqrt{\frac{T_w}{T_i}} \left[ \frac{\gamma \sqrt{\pi}}{s}(1+\text{erf}(\gamma s))+ \frac{1}{s^2}e^{-\gamma^2s^2} \right],
\end{equation}
where $T_w$ is the wall temperature of the satellite and $T_i$ is the temperature of the incident particles \cite{Hild.2021}.
Finally, the momentum components of the reflected particle have to be expressed. Therefore, Eq. \ref{eq:at_impulse} can be transformed into 
\begin{equation}
\label{eq:I_r}
I_r = \sqrt{I_i^2-\alpha_T(I_i^2-I_w^2)}.
\end{equation}
Using Eq.~\ref{eq:impulse} on $I_i$ leads to
\begin{equation}
I_i = \sqrt{p_i^2+\tau_i^2},
\end{equation}
which now can be used to express Eq.~\ref{eq:I_r} as
\begin{equation}
I_r = \sqrt{p_i^2+\tau_i^2-\alpha_T(p_i^2+\tau_i^2-p_w^2)}.
\end{equation}
Inserting the above derived parameters into the following equations from Hild for $\tau_r$ and $p_r$ \cite{Hild.2021}
\begin{equation}
\tau_r = (1-g)\tau_i \frac{I_r}{I_i}
\end{equation} 
\begin{equation}
p_r = g \cdot I_r + (1-g)\cdot p_i \frac{I_r}{I_i},
\end{equation}
the tangential and normal momentum carried away from the wall by the particle can be described using the parameters $\alpha_T$ and $g$:
\begin{equation}
\tau_r = (1-g)\tau_i \sqrt{\frac{p_i^2+\tau_i^2-\alpha_T (p_i^2+\tau_i^2-p_w^2)}{p_i^2+\tau_i^2}},
\end{equation} 
\begin{multline}
p_r = \sqrt{p_i^2+\tau_i^2-\alpha_T(p_i^2+\tau_i^2-p_w^2)} \\
\cdot \left( g+(1-g) \frac{p_i}{\sqrt{p_i^2+\tau_i^2}} \right),
\end{multline}
and thus, $\alpha_n$ and $\sigma_t$ can be obtained utilizing Eq.~\ref{eq:sigma_t} and Eq.~\ref{eq:alpha_n} for given $\alpha_T$ and $g$. However, it should be noted that $\tau_i$, $p_i$ and $p_w$ are determined using Sentman's equation in a 2D simplified version, which is valid for one area element with a given orientation regarding the velocity vector. Hence, the values for $\tau_i$, $p_i$ and $p_w$ as well as $\tau_r$ and $p_r$ change with different orientation of the considered area element. Assuming a constant energy and momentum accommodation over the satellite, the respective parameters may still be used on one representative area element in order to derive the normal energy accommodation coefficient and tangential momentum accommodation coefficient. 

The derived input parameters for the use of the CLL model according to the given $\alpha_T$ in Tab.~\ref{tab:surface_properties} can be found in Tab.~\ref{tab:CLL_input} and were used for all the CLL ADBSat calculations within this work. 
\begin{table}
	\centering
		\begin{tabular}{l|ll} 
		$\alpha_T$ [-] & $\alpha_n$ [-] & $\sigma_t$ [-]\\ \hline 
		0 & 0 & 0 \\ 
		0.09  & 0.09 & 0.0459 \\ 
		0.30 & 0.30 & 0.1627 \\ 
	\end{tabular}
	\caption{Derived CLL input parameters according to the specified $\alpha_T$ gradation in Tab.~\ref{tab:surface_properties}.}
	\label{tab:CLL_input}%
\end{table}

 \bibliographystyle{elsarticle-num} 
 \bibliography{cas-refs}





\end{document}